\documentclass[final,conference, twoside, 10pt, letterpaper]{IEEEtran}
\usepackage{Bamble}
\usepackage{Bmacros}
\usepackage[noadjust]{cite}
\usepackage{accents}
\usepackage{balance}

\newcommand{\myfloor}[1]{\lfloor #1 \rfloor}
\newcommand{\myceil}[1]{\lceil #1 \rceil}
\newcommand{\period}{\mathsf{p}}
\newcommand{\qeriod}{\mathsf{q}}
\newcommand{\deriod}{\mathsf{d}}

\newcommand{\mixing}{M}

\newcommand{\twobibs}[2]{#2} 

\begin{document}
\IEEEoverridecommandlockouts \title{Constant Weight Polar Codes through Periodic Markov Processes}
\date{}
\author{\IEEEauthorblockN{Boaz~Shuval, Ido~Tal\\
The Andrew and Erna Viterbi Faculty of Electrical and Computer Engineering,\\
Technion, Haifa 32000, Israel.\\
Email: \{\texttt{bshuval@}, \texttt{idotal@ee.}\}\texttt{technion.ac.il}}
}

\maketitle

\begin{abstract}
	Constant weight codes can arise from an input process sampled from a periodic Markov chain. A previous result showed that, in general, polarization does \emph{not} occur for input-output processes with an underlying \emph{periodic} Markov chain. In this work, we show that if we fix the initial state of an underlying periodic Markov chain, polarization \emph{does} occur. Fixing the initial state is aligned with ensuring a constant weight code.
\end{abstract}

\section{Introduction}
Polar codes have been shown to achieve the capacity or information rate of many channel models \cite{Arikan:09p, Arikan:10c, STA:09a, SasogluTal:19p, ShuvalTal:19.2p, TPFV:22p, PfisterTal:21c,AravaTal:23c, MahdavifarVardy:11p, Mahdavifar:20p, HofShamai:10a, Hof+:13p}. Specifically, in \cite{SasogluTal:19p} and \cite{ShuvalTal:19.2p}, polar codes were shown to achieve the information rate of input-output processes with an underlying finite-state, \emph{aperiodic}, and irreducible Markov (FAIM) process. Moreover, in \cite[Theorem 4]{SasogluTal:19p} it was shown by a counter-example that aperiodicity is a necessary condition. In this paper, we prove that polarization does occur for periodic processes, provided that the initial state is fixed. This setting is natural in the context of codewords constrained to be of a constant weight.

\begin{figure}
	\begin{center}
		\begin{tikzpicture}[>=latex]
			\node[circle, minimum size = 0.8cm, inner sep = 0, outer sep = 0, draw] (N0) {$0$};
			\node[circle, minimum size = 0.8cm, inner sep = 0, outer sep = 0, draw, on grid, right = 1.3of N0] (N1) {$1$};
			\node[circle, minimum size = 0.8cm, inner sep = 0, outer sep = 0, draw, on grid, right = 1.3of N1] (N2) {$2$};
			\node[on grid, right = 1 of N2] (N2x){};
\node[circle, minimum size = 0.8cm, inner sep = 0, outer sep = 0, draw, on grid, right = 3 of N2] (N5) {$b-2$};
			\node[on grid, left = 1 of N5] (N5x){};
			\node[circle, minimum size = 0.8cm, inner sep = 0, outer sep = 0, draw, on grid, right = 1.3of N5] (N6) {$b-1$};
			\node[circle,fill,minimum size = 0.1cm, inner sep = 0, outer sep = 0, draw] at ($(N2x)!0.5!(N5x)$) (N25) {}; 
			\node[circle,fill,minimum size = 0.1cm, inner sep = 0, outer sep = 0, draw, on grid, left = 0.3 of N25] {}; 
			\node[circle,fill,minimum size = 0.1cm, inner sep = 0, outer sep = 0, draw, on grid, right = 0.3 of N25] {}; 

			\draw[->,red] (N0) -> (N1); 
			\draw[->,red] (N1) -> (N2); 
\draw[->,red] (N2) -> (N2x); 
			\draw[->,red] (N5x) -> (N5); 
			\draw[->,red] (N5) -> (N6); 

			\draw[->,red] (N6) to[in = 50 , out = 130, looseness=0.6] (N0);

			\draw[->,blue] (N0) to[in = 250 , out = 290, min distance = 1cm, looseness = 5]  (N0); 
			\draw[->,blue] (N1) to[in = 250 , out = 290, min distance = 1cm, looseness = 5]  (N1); 
			\draw[->,blue] (N2) to[in = 250 , out = 290, min distance = 1cm, looseness = 5]  (N2); 
\draw[->,blue] (N5) to[in = 250 , out = 290, min distance = 1cm, looseness = 5]  (N5); 
			\draw[->,blue] (N6) to[in = 250 , out = 290, min distance = 1cm, looseness = 5]  (N6); 
		\end{tikzpicture}
	\end{center}
	\caption{Markov chain for producing sequences whose weight modulo $b$ is constrained to some integer $0 \leq a < b$. Blue and red arrows are labelled `\textcolor{blue}{$0$}' and `\textcolor{red}{$1$}', respectively. All edge probabilities are $1/2$. The chain is aperiodic. To construct the sequences, we constrain the initial state to `$0$' and the final state to `$a$'.}
	\label{fig: chain for weight a mod b}
\end{figure}
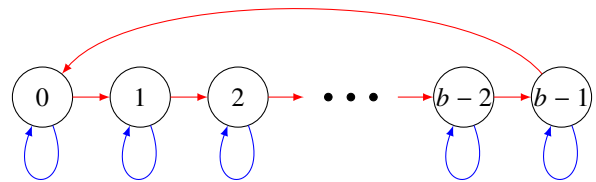

An example of an irreducible Markov chain of period $\period=4$ for producing sequences of constant weight $N/2$ (i.e., normalized weight $1/2$) is given in \Cref{fig: weight half Markov chain}. This holds if the initial state is $\varepsilon$ and the sequence length $N$ is a multiple of $4$. A few things to note:
\begin{itemize}
	\item We produce only a subset of the sequences of weight $N/2$.
	\item All the sequences of length $4$ have equal probability, $1/6$, where $6 = \binom{4}{2}$ is the number of sequences of length $4$ and weight $2$.
	\item The number of states can be drastically reduced, while producing the same set of sequences at the same probabilities, see \Cref{fig: condensed weight half Markov chain}.
	\item Enlarging the period $\period$ from $4$ to a larger number will allow us to produce a larger subset of sequences of weight $N/2$, at the expense of more states.
	\item If the sequence length $N$ is not a multiple of the period $\period$, we must also constrain the final state to be in a certain set. For example, in \Cref{fig: weight half Markov chain} and $N=6$, the final state must be either `$10$' or `$01$'. That is, of weight $1$.
	\item For $\alpha = a/b$ rational, a similar graph can be constructed for sequences of weight $\myfloor{\alpha N}$, when $N$ is not a multiple of $b$. The same holds for weight $\myceil{\alpha N}$. This again requires constraining the final state to be of a certain weight.
	\item For $\alpha \leq \beta$ rational, a similar graph can be constructed for sequences of weight  between $\alpha N$ and $\beta N$. Now, generally, the final state is constrained to have a range of weights.
	\item For positive integers $a$ and $b$, the aperiodic graph in \Cref{fig: chain for weight a mod b} can ensure sequences whose weight modulo $b$ is $a$. This is obtained by fixing the initial state to be `$0$' and the final state to be `$a$'.
\end{itemize}
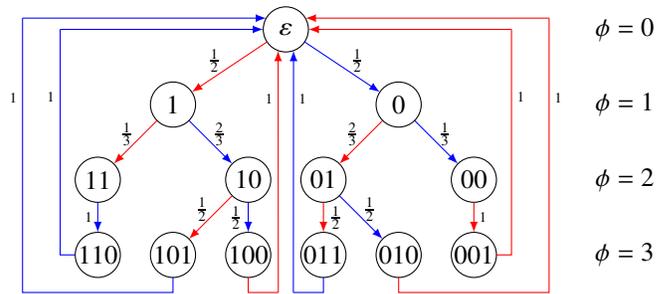
\begin{figure}
	\begin{center}
		\begin{tikzpicture}[>=latex]
				
			\node[circle, minimum size = 0.6cm, inner sep = 0, outer sep = 0, draw] (epsilon) {$\varepsilon$};
			\node[circle, minimum size = 0.6cm, inner sep = 0, outer sep = 0, draw, on grid, below right = 1 and 1.5 of epsilon] (N0) {$0$};
			\node[circle, minimum size = 0.6cm, inner sep = 0, outer sep = 0, draw, on grid, below left = 1 and 1.5  of epsilon] (N1) {$1$};
			\node[circle, minimum size = 0.6cm, inner sep = 0, outer sep = 0, draw, on grid, below right = of N0] (N00) {$00$};
			\node[circle, minimum size = 0.6cm, inner sep = 0, outer sep = 0, draw, on grid, below left = of N0] (N01) {$01$};
			\node[circle, minimum size = 0.6cm, inner sep = 0, outer sep = 0, draw, on grid, below = of N00] (N001) {$001$};
			\node[circle, minimum size = 0.6cm, inner sep = 0, outer sep = 0, draw, on grid, below right = of N01] (N010) {$010$};
			\node[circle, minimum size = 0.6cm, inner sep = 0, outer sep = 0, draw, on grid, below = of N01] (N011) {$011$};
\node[circle, minimum size = 0.6cm, inner sep = 0, outer sep = 0, draw, on grid, below left = of N1] (N11) {$11$};
			\node[circle, minimum size = 0.6cm, inner sep = 0, outer sep = 0, draw, on grid, below right  = of N1] (N10) {$10$};
			\node[circle, minimum size = 0.6cm, inner sep = 0, outer sep = 0, draw, on grid, below = of N11] (N110) {$110$};
			\node[circle, minimum size = 0.6cm, inner sep = 0, outer sep = 0, draw, on grid, below = of N10] (N100) {$100$};
			\node[circle, minimum size = 0.6cm, inner sep = 0, outer sep = 0, draw, on grid, below left = of N10] (N101) {$101$};

			\draw[->,blue] (epsilon) -> node[pos =0.7, above = -0.05 cm, black] {\tiny $\frac{1}{2}$} (N0) ;
			\draw[->,blue] (N0) ->  node[pos =0.7, above = -0.05 cm, black] {\tiny $\frac{1}{3}$} (N00);
			\draw[->,blue] (N01) -> node[pos =0.7, above = -0.05 cm, black] {\tiny $\frac{1}{2}$} (N010);
			\draw[->,blue] (N1) -> node[pos =0.7, above = -0.05 cm, black] {\tiny $\frac{2}{3}$} (N10);
			\draw[->,blue] (N11) -> node[pos =0.5, left = -0.05 cm, black] {\tiny $1$} (N110);
			\draw[->,blue] (N10) -> node[pos =0.5, left = -0.05 cm, black] {\tiny $\frac{1}{2}$} (N100);
			
			\draw[->,red] (epsilon) ->node[pos =0.7, above = -0.05 cm, black] {\tiny $\frac{1}{2}$} (N1);
			\draw[->,red] (N1) -> node[pos =0.7, above = -0.05 cm, black] {\tiny $\frac{1}{3}$}(N11);
			\draw[->,red] (N0) -> node[pos =0.7, above = -0.05 cm, black] {\tiny $\frac{2}{3}$} (N01);
			\draw[->,red] (N10) -> node[pos =0.7, above = -0.05 cm, black] {\tiny $\frac{1}{2}$} (N101);
			\draw[->,red] (N01) -> node[pos =0.5, right = -0.05 cm, black] {\tiny $\frac{1}{2}$} (N011);
			\draw[->,red] (N00) -> node[pos =0.5, right = -0.05 cm, black] {\tiny $1$} (N001);

			\node[on grid, left = 0.5 of N110] (N110left) {}; 
			\node[on grid, above = 3 of N110left] (epsleft) {}; 
			\draw[blue] (N110.west) -- (N110left.center) -- node[pos =0.7, left = -0.05 cm, black] {\tiny $1$} (epsleft.center);  
			\draw[blue] (epsleft.center) edge[->] (epsilon.west);

			\node[on grid, below = 0.5 of N101] (N101below) {}; 
			\node[on grid, left = 2 of N101below] (N101belowleft) {}; 
			\node[on grid, above left = 0.15 and 0.5 of epsleft] (epsleftleft) {}; 
			\draw[blue] (N101.south) -- (N101below.center) --(N101belowleft.center) --node[pos =0.71, left = -0.05 cm, black] {\tiny $1$}   (epsleftleft.center);
			\draw[blue] (epsleftleft.center) edge[->] (epsilon.150);

			\node[on grid, below = 0.5 of N100] (N100below) {}; 
			\node[on grid, right = 0.40 of N100below] (epsbelowleft) {}; 
			\draw[red] (N100.south) -- (N100below.center) -- (epsbelowleft.center) edge[->] node[pos =0.8, left = -0.05 cm, black] {\tiny $1$} (epsilon.250);

			\node[on grid, right = 0.5 of N001] (N001right) {}; 
			\node[on grid, above = 3 of N001right] (epsright) {}; 
			\draw[red] (N001.east) -- (N001right.center) --node[pos =0.7, right = -0.05 cm, black] {\tiny $1$} (epsright.center);
			\draw[red] (epsright.center) edge[->] (epsilon.east);

			\node[on grid, below = 0.5 of N010] (N010below) {}; 
			\node[on grid, right = 2 of N010below] (N010belowright) {}; 
			\node[on grid, above right = 0.15 and 0.5 of epsright] (epsrightright) {}; 
			\draw[red] (N010.south) -- (N010below.center) -- (N010belowright.center) --node[pos =0.71, right = -0.05 cm, black] {\tiny $1$} (epsrightright.center);
			\draw[red] (epsrightright.center) edge[->] (epsilon.30);
                        
			\node[on grid, below = 0.5 of N011] (N011below) {}; 
			\node[on grid, left = 0.40 of N011below] (epsbelowright) {}; 
			\draw[blue] (N011.south) -- (N011below.center) -- (epsbelowright.center) edge[->] node[pos =0.8, right = -0.05 cm, black] {\tiny $1$}(epsilon.290);

			\node[on grid, right = 4.5 of epsilon] (phi) {$\phi = 0$}; 
			\node[on grid, below = 1 of phi] {$\phi = 1$}; 
			\node[on grid, below = 2 of phi] {$\phi = 2$}; 
			\node[on grid, below = 3 of phi] {$\phi = 3$};

		\end{tikzpicture}
	\end{center}
	\caption{Markov chain for producing sequences of weight $N/2$. Labels for blue and red arrows are  `\textcolor{blue}{$0$}' and `\textcolor{red}{$1$}', respectively. Transition probabilities between states are shown on the arrows. The state $\varepsilon$ is the start and end state, at which the output is ensured to have weight $N/2$. The chain has period $\period=4$; each horizontal layer of the graph is a different phase $\phi$.}
\label{fig: weight half Markov chain}
\end{figure}

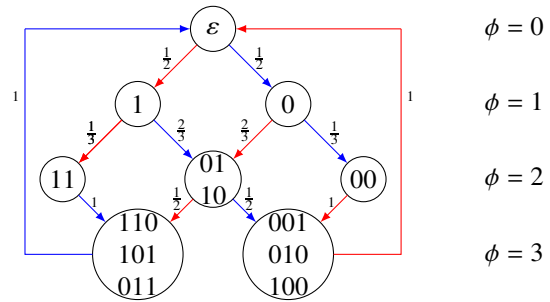
\begin{figure}
	\begin{center}
		\begin{tikzpicture}[>=latex]
				
			\node[circle, minimum size = 0.6cm, inner sep = 0, outer sep = 0, draw] (epsilon) {$\varepsilon$};
			\node[circle, minimum size = 0.6cm, inner sep = 0, outer sep = 0, draw, on grid, below right = 1 and 1 of epsilon] (N0) {$0$};
			\node[circle, minimum size = 0.6cm, inner sep = 0, outer sep = 0, draw, on grid, below left = 1 and 1  of epsilon] (N1) {$1$};
			\node[circle, minimum size = 0.6cm, inner sep = 0, outer sep = 0, draw, on grid, below right = of N0] (N00) {$00$};
			\node[circle, minimum size = 0.6cm, inner sep = 0, outer sep = 0, draw, on grid, below left = of N1] (N11) {$11$};
			\node[circle, minimum size = 0.6cm, inner sep = 0, outer sep = 0, draw, on grid, below = 2 of epsilon, align = center] (N01N10) {$01$\\$10$};
			\node[circle, inner sep = 0, outer sep = 0, draw, on grid, below = 2 of N1, align = center] (N110N101N011) {$110$\\$101$ \\$011$};
			\node[circle, inner sep = 0, outer sep = 0, draw, on grid, below = 2 of N0, align = center] (N001N010N100) {$001$\\ $010$ \\$100$};

			\draw[->,blue] (epsilon) -> node[pos =0.7, above = -0.05 cm, black] {\tiny $\frac{1}{2}$} (N0) ;
			\draw[->,red] (epsilon) ->node[pos =0.7, above = -0.05 cm, black] {\tiny $\frac{1}{2}$} (N1);
			\draw[->,red] (N1) -> node[pos =0.7, above = -0.05 cm, black] {\tiny $\frac{1}{3}$}(N11);
			\draw[->,blue] (N0) ->  node[pos =0.7, above = -0.05 cm, black] {\tiny $\frac{1}{3}$} (N00);
			\draw[->,blue] (N1) -> node[pos =0.7, above = -0.05 cm, black] {\tiny $\frac{2}{3}$} (N01N10);
			\draw[->,red] (N0) -> node[pos =0.7, above = -0.05 cm, black] {\tiny $\frac{2}{3}$} (N01N10);
			\draw[->,red] (N1) -> node[pos =0.7, above = -0.05 cm, black] {\tiny $\frac{1}{3}$}(N11);
			\draw[->,blue] (N11) -> node[pos =0.25, right = -0.05 cm, black] {\tiny $1$} (N110N101N011);
			\draw[->,red] (N00) -> node[pos =0.25, left = -0.05 cm, black] {\tiny $1$} (N001N010N100);
			\draw[->,blue] (N01N10) -> node[pos =0.7, above = -0.05 cm, black] {\tiny $\frac{1}{2}$} (N001N010N100);
			\draw[->,red] (N01N10) -> node[pos =0.7, above = -0.05 cm, black] {\tiny $\frac{1}{2}$} (N110N101N011);
                        
			\node[on grid, left = 1.5 of N110N101N011] (N110left) {}; 
			\node[on grid, above = 3 of N110left] (epsleft) {}; 
			\draw[blue] (N110N101N011.west) -- (N110left.center) -- node[pos =0.7, left = -0.05 cm, black] {\tiny $1$} (epsleft.center);  
			\draw[blue] (epsleft.center) edge[->] (epsilon.west);

			\node[on grid, right = 1.5 of N001N010N100] (N001right) {}; 
			\node[on grid, above = 3 of N001right] (epsright) {}; 
			\draw[red] (N001N010N100.east) -- (N001right.center) -- node[pos =0.7, right = -0.05 cm, black] {\tiny $1$} (epsright.center);  
			\draw[red] (epsright.center) edge[->] (epsilon.east);

			\node[on grid, right = 4 of epsilon] (phi) {$\phi = 0$}; 
			\node[on grid, below = 1 of phi] {$\phi = 1$}; 
			\node[on grid, below = 2 of phi] {$\phi = 2$}; 
			\node[on grid, below = 3 of phi] {$\phi = 3$};

		\end{tikzpicture}
	\end{center}
	\caption{Condensed Markov chain for producing sequences of weight $N/2$. Here, states from \Cref{fig: weight half Markov chain} of the same phase and weight are merged. The probability of a blue (red) edge from a node with phase $\phi$ and weight $w$ to a node with phase $\phi+1$ and weight $w$ ($w+1$) is the number of valid sequences, i.e. length $4$ and weight $2$, whose weight in the prefix of length $\phi+1$ is $w$ ($w+1$) divided by the number of valid sequences whose weight in the prefix of length $\phi$ is $w$. } \label{fig: condensed weight half Markov chain}
\end{figure}

\section{Setting and Main Theorem}\label{sec:setting}
Although we will ultimately fix the initial state, we start by considering a strictly stationary process, $(X_j,Y_j, S_j)_{j\in \mathbb{Z}}$. Here, $X_j$ is binary, $Y_j \in \mathcal{Y}$, and $S_j \in \mathcal{S}$, with $\mathcal{Y}$ and $\mathcal{S}$ finite alphabets. We call $S_j$ the state at time $j$. The conditional distribution of the process is $\prrv{X_j,Y_j,S_j|S_{j-1}}{}$, independent of $j$ by stationarity. The distribution has a Markov property: conditioned on $S_{j-1}$, the random variables $X_k,Y_k,S_k$ are independent of $X_l,Y_l, S_{l-1}$ for any $l < j \leq k$.

The state sequence $S_j$ is a finite-state irreducible Markov (FIM) process. By the Perron-Frobenius theorem \cite[Theorem 8.4.4 and Corollary 8.1.30]{HornJohnson:85b}, it indeed has a unique stationary distribution $\pi$, which is positive. That is, for any state $s \in \mathcal{S}$, $\pi(s) > 0$. We denote the period of the state sequence by $\period$. That is, we can partition $\mathcal{S}$ into $\period$ classes, which we call \emph{phases}. The phase of $s \in \mathcal{S}$ is denoted $\phi(s)$, an integer between $0$ and $\period-1$. This partition is such that $\Prob{S_{j+1} = s' | S_j = s}$ is positive only if $\phi(s') \equiv \phi(s) + 1 \pmod \period$. For example, in \Cref{fig: weight half Markov chain}, each horizontal layer corresponds to a different phase.

The conditional entropy, Bhattacharyya, and total variation parameters are key to proving polarization theorems. For a binary random variable $U$, general random variables $W$, $Q$, and an event $A$,
\begin{IEEEeqnarray*}{rCl}
	H(W|Q,A) & = & -\sum_{w,q} \prrv{W,Q|A}{w,q} \log_2 \prrv{W|Q,A}{w|q}, \\
	Z(U|Q,A) & = & 2 \sum_q \sqrt{\prrv{U,Q|A}{0,q}\prrv{U,Q|A}{1,q}}, \\
	K(U|Q,A) & = & \sum_q \left|\prrv{U,Q|A}{0,q}-\prrv{U,Q|A}{1,q}\right|.
\end{IEEEeqnarray*}

The entropy rate of our process, excluding the states, is
\begin{equation} \label{eq:HXY as limit}
\mathcal{H}_{X|Y} = \lim_{N\to\infty} \frac1N H(X_1^N|Y_1^N).
\end{equation}
We also denote
\begin{equation} \label{eq:HX as limit}
\mathcal{H}_{X} = \lim_{N\to\infty} \frac1N H(X_1^N).
\end{equation}
These limits exist, see~\cite[Section II]{SasogluTal:19p}.

Following is our main theorem. By the Honda-Yamamoto scheme \cite{HondaYamamoto:12p}, it implies that a polar coding scheme with rate approaching $\mathcal{H}_{X} - \mathcal{H}_{X|Y}$ and vanishing error probability exists, if we condition on either the initial state or the initial state and the set of final states. 

\begin{theorem}
	\label{thm:main}
	Let $(X_j, Y_j, S_j)_{j \in \mathbb{Z}}$ be a FIM process, where $(S_j)_{j \in \mathbb{Z}}$ has period $\period$. For $N = 2^n$, denote by $U_1^N$ the polar transform of $X_1^N$, and fix $0 < \beta <1/2$. Fix $0 \leq \varphi < \period$, and a non-empty set $\Psi_0 \subseteq \mathcal{S}$ such that all states $s_0 \in \Psi_0$ have phase $\phi(s_0) = \varphi$.  Furthermore, let $\Psi_N$ be a  non-empty subset of $\mathcal{S}$ such that all states $s_N \in \Psi_N$ have phase $\phi(s_N) = \varphi + N \bmod{\period}$. 
Denote the event $A_N$ as
	\[
		A_N \triangleq \{S_0 \in \Psi_0 \quad \mbox{and} \quad S_N \in \Psi_N \} .
	\]Then,
	\begin{IEEEeqnarray*}{rCl}
	\lim_{n \to \infty} \frac{1}{N}\left|\left\{ i : Z(U_i|U_1^{i-1},Y_1^N, A_N ) < 2^{-N^\beta} \right\} \right| & = & 1 - \mathcal{H}_{X|Y} , \\
	\lim_{n \to \infty} \frac{1}{N}\left|\left\{ i : K(U_i|U_1^{i-1}, A_N) < 2^{-N^\beta} \right\} \right| & = & \mathcal{H}_{X} ,
	\end{IEEEeqnarray*}
\end{theorem}

We note that the encoding and decoding complexity of our coding scheme is $O(|\mathcal{S}|^3 N \log N)$, utilizing successive-cancellation trellis decoding
\cite{Wang+:15c,HondaYamamoto:16a} and the Honda-Yamamoto scheme \cite{HondaYamamoto:12p}. The only difference is that rather than marginalizing $S_0, S_N$ over all states in $|\mathcal{S}|$, we marginalize $S_0$ only over $\Psi_0$ and $S_N$ only over $\Psi_N$.

Throughout this paper, we assume the setting of \Cref{thm:main}. We denote by $U_1^N$, $V_1^N$ the polar transforms of $X_1^N$ and $X_{N+1}^{2N}$, respectively. We further denote $Q_i = (U_1^{i-1}, Y_1^N)$ and $R_i = (V_1^{i-1}, Y_{N+1}^{2N})$, for $1 \leq i \leq N$.

Our proof of \Cref{thm:main} will follow by first showing slow polarization and then showing fast polarization. We factor the period $\period$ into 
\begin{equation} \label{eq:period factorization} 
	\period = \deriod \cdot \qeriod, 
\end{equation}
where $\qeriod$ is odd and $\deriod$ is a power of $2$. In essence, we show that both slow and fast polarization occur if we fix the  phase modulo $\deriod$ of the initial state. Our proofs are extensions of the ideas presented in \cite{Arikan:09p}, \cite{Sasoglu:12b}, \cite{SasogluTal:19p},\cite{ShuvalTal:19.2p}, and \cite{TPFV:22p}, while also taking into account the above.

For  $0 \leq \Delta < \deriod$, define
\[ \mathcal{S}(\Delta) = \{ s \in \mathcal{S} : \phi(s) \equiv \Delta \pmod{\deriod} \}.\]
We denote the event
	\[
		D(\Delta) \triangleq \{S_0 \in \mathcal{S}(\Delta) \} =  \left\{\phi(S_0) \equiv \Delta \pmod \deriod \right\} .
	\]
	Since $N = 2^n$ and $\deriod$ are both powers of $2$, we have
	\begin{equation}
		\label{eq:DDelta alternate}
		D(\Delta) = \{S_N \in \mathcal{S}(\Delta)\} , \quad \mbox{for $N \geq \deriod$}.  
\end{equation}
Observe that by stationarity, we have for $0\leq \Delta \leq \deriod$ 
\begin{equation} \label{eq:prob of DDelta} \Prob{D(\Delta)} = 1/\deriod. 
\end{equation}

For $N = 2^n \geq \deriod$ denote 
\begin{multline}
	\label{eq:SN2}
	\mathcal{S}_N^2(\Delta) = \big\{ (s_0,s_N) \in \mathcal{S}^2 : s_0 \in \mathcal{S}(\Delta) \quad \mbox{and} \\ \phi(s_N) - \phi(s_0) \equiv N \pmod \period \big\} ,
\end{multline}
and
	\begin{multline}
		\label{eq:correct phase differences and dhase}
\mathcal{S}^3_N(\Delta) = \big\{ (s_0,s_N,s_{2N}) \in \mathcal{S}^3 : (s_0,s_N) \in \mathcal{S}_N^2(\Delta) \quad \mbox{and} \\ (s_N,s_{2N}) \in \mathcal{S}_N^2(\Delta)\big\} .
	\end{multline}
Observe that for $N = 2^n \geq \deriod$, 
	\begin{IEEEeqnarray*}{rCl}
D(\Delta) \; \mbox{holds} &\Leftrightarrow& \!\{S_0 \in \mathcal{S}(\Delta) \} \Leftrightarrow \{S_N \in \mathcal{S}(\Delta)\}\Leftrightarrow \{S_{2N} \in \mathcal{S}(\Delta) \} \\ 
					  &\Leftrightarrow&\!\{(S_0, S_N)\! \in \!\mathcal{S}_N^2(\Delta)\} \!\Leftrightarrow \!\{(S_0, S_N, S_{2N}) \!\in \!\mathcal{S}_N^3(\Delta)\}. 
	 \end{IEEEeqnarray*}

The following definition and the observation and lemma that follow will be of use in the sequel. 
\begin{definition}\label{def:simeq}
Denote $f(N) \simeq g(N)$ if $f(N) - g(N)$ can be upper and lower bounded by expressions independent of $N$. It is easily shown that $\simeq$ is a transitive relation. 
\end{definition}
Observe that if $f(N) \simeq g(N)$ and the limit $\lim_{N \to \infty} f(N)/N$ exists, then $\lim_{N \to \infty} g(N)/N$ exists, and the limits coincide. 
\begin{lemma}\label{lemm:HVWQA simeq}
If $V$, $Q$, and $W$ are random variables, $A$ is an event, and the support of $W$ is finite, then  
	\begin{equation}
		\label{eq:technical simeq relations}
		H(V,W|Q,A) \simeq  H(V|Q,A) \simeq H(V|Q,W,A) , 
	\end{equation}
\end{lemma}
\begin{IEEEproof}
 This is by the chain rule. For the left $\simeq$, by the chain rule, $H(V,W|Q,A) = H(V|Q,A)+H(W|V,Q,A)$. Since the support of $W$ is finite, the claim follows.  For the right $\simeq$, we use the chain rule on $H(W,V|Q,A)$ both ways and obtain
\[
H(V|W,Q,A) - H(V|Q,A) = H(W|V,Q,A) - H(W|Q,A).
	\]
	Since both terms on the right-hand side are bounded between $0$ and the base-2 logarithm of the support of $W$, we conclude that $H(V|Q,W,A) \simeq H(V|Q,A)$. 	
\end{IEEEproof}

Finally, as a shorthand, for random variables $X,Y$ we denote $P_{X|Y}(x|y) = \Prob{X=x|Y=y}$. The extension for multiple random variables is immediate. 

\section{Proof of Slow Polarization}
The proof of slow polarization comprises two steps: showing a convergence to a limiting random variable and then showing that this random variable is either $0$ or $1$, with probability $1$. 

For a given $0 \leq \Delta  < \deriod$, $N=2^n$, and $1 \leq i \leq N$, denote
\begin{IEEEeqnarray}{rCl}
	\hat{H}_n(\Delta) &=& H(U_i|Q_i, S_0, S_N, D(\Delta)), 
	\label{eq:HnDelta} \\ 
	\hat{H}_n^-(\Delta) &=& H(U_i+V_i|Q_i, R_i, S_0, S_{2N}, D(\Delta)). \IEEEeqnarraynumspace
	\label{eq:HnMinusDelta} \\ 
	H_n(\Delta) &=& H(U_i|Q_i, D(\Delta)). \label{eq:HnBarDelta}
\end{IEEEeqnarray}
\begin{lemma}
	\label{lemm:H_n converges}
	Fix $0 \leq \Delta  < \deriod$. Consider the random process
	\[
J_n = 1 + \left\langle B_0,B_1,\ldots,B_n \right\rangle_2  = 1 + \sum_{j = 0}^{n-1} B_j 2^{n-j} ,
	\]
	where the $B_j$ are i.i.d.\ Bernoulli $1/2$ random variables.
Then, the random processes $(\hat{H}_n(\Delta))_{n=1}^\infty$ and $(H_n(\Delta))_{n=1}^\infty$ where $i = J_n$ converge a.s., in $L_1$, and in probability to limiting random variables $\hat{H}_\infty(\Delta) \in [0,1]$ and $H_\infty(\Delta) \in [0,1]$, respectively.
\end{lemma}

\begin{IEEEproof}
Take $n_0$ large enough such that $2^{n_0} \geq \deriod$. The proof follows by showing that the processes $(\hat{H}_n(\Delta))_{n=n_0}^\infty$ and $(H_n(\Delta))_{n=n_0}^\infty$ are a bounded sub-martingale and a bounded a super-martingale, respectively, with respect to $(J_n)_{n=1}^{\infty}$, see \cite[Theorems 7.3.1 and 14.2.1 and Proposition 5.2.3]{Rosenthal:06b}. Since we are considering entropies of binary random variables, both $\hat{H}_n(\Delta)$ and $H_n(\Delta)$  are clearly bounded between $0$ and $1$. It remains to show that $\Exp{\hat{H}_{n+1}(\Delta)|J_1, J_2,\ldots,J_n} \geq \hat{H}_n(\Delta)$ and $\Exp{H_{n+1}(\Delta)|J_1, J_2,\ldots,J_n} \leq H_n(\Delta)$, for $n \geq n_0$. Indeed, let $i=J_n$ be the random index corresponding to $\hat{H}_n(\Delta)$. Then, \begin{IEEEeqnarray*}{rCl}
		\IEEEeqnarraymulticol{3}{l}{\Exp{\hat{H}_{n+1}(\Delta) | J_1, J_2,\ldots, J_n}} \\
		\quad &\eqann{a}&  \frac{1}{2} H(U_i + V_i| Q_i, R_i, S_0, S_{2N}, D(\Delta)) \\
				    && {} + \frac{1}{2} H(V_i| U_i + V_i, Q_i, R_i, S_0, S_{2N}, D(\Delta) ) \\
				    &\eqann{b}&\frac{1}{2} H(U_i , V_i| Q_i, R_i, S_0, S_{2N}, D(\Delta) )  \\
				    & \eqann[\geq]{c} &\frac{1}{2} H(U_i , V_i| Q_i, R_i, S_0, S_N, S_{2N}, D(\Delta) )  \\
				    & \eqann{d} &\frac{1}{2} H(U_i | Q_i, R_i, S_0, S_N, S_{2N}, D(\Delta) )  \\
				    & & {} + \frac{1}{2} H(V_i | U_i, Q_i, R_i, S_0, S_N, S_{2N}, D(\Delta) )  \\
				    & \eqann{e} &\frac{1}{2} H(U_i | Q_i, S_0, S_N, D(\Delta) )  
				     + \frac{1}{2} H(V_i | R_i, S_N, S_{2N}, D(\Delta) )  \\
				    & \eqann{f} & \hat{H}_n(\Delta) .
\end{IEEEeqnarray*}
	Above, \eqannref{a} is since $B_{n+1}$ is Bernoulli $1/2$, \eqannref{b} is by the chain rule, \eqannref{c} is since conditioning reduces entropy, \eqannref{d} is again by the chain rule, \eqannref{e} is by the Markov property and~\eqref{eq:DDelta alternate}, and \eqannref{f} is by stationarity. 
	
	For $H_n(\Delta)$ we have
	\begin{IEEEeqnarray*}{rCl}
		\IEEEeqnarraymulticol{3}{l}{\Exp{H_{n+1}(\Delta) | J_1, J_2,\ldots, J_n}} \\
		\quad &\eqann{a}&  \frac{1}{2} \Big(H(U_i + V_i| Q_i, R_i, D(\Delta))  + (V_i| U_i + V_i, Q_i, R_i, D(\Delta) )  \Big) \\ 
				    &\eqann{b}&\frac{1}{2} H(U_i , V_i| Q_i, R_i, D(\Delta) )  \\
				    & \eqann{c} &\frac{1}{2} H(U_i | Q_i, R_i, D(\Delta) )  + \frac{1}{2} H(V_i | U_i, Q_i, R_i, D(\Delta) )  \\
				    & \eqann[\leq]{d} &\frac{1}{2} H(U_i | Q_i, D(\Delta) ) + \frac{1}{2} H(V_i | R_i, D(\Delta) )  \\
				    & \eqann{e} & H_n(\Delta) .
\end{IEEEeqnarray*}
	Above, \eqannref{a} is since $B_{n+1}$ is Bernoulli $1/2$, \eqannref{b} is by the chain rule, \eqannref{c} is again by the chain rule, \eqannref{d} is since conditioning reduces entropy and~\eqref{eq:DDelta alternate}, and \eqannref{e} is by stationarity. 
\end{IEEEproof}

For $N > 0$, denote 
	\begin{multline}
		\label{eq:correct phase differences}
		\mathcal{S}_N^3 = \big\{ (s_0,s_N,s_{2N}) \in \mathcal{S}^3 : \\ \phi(s_N) - \phi(s_0) \equiv \phi(s_{2N}) - \phi(s_N) \equiv N \pmod \period \big\} .
	\end{multline}
	Further denote, for fixed $N$ clear from the context,
	\begin{equation} \label{eq:p s0 sN s2N}
		\pi(s_0, s_N, s_{2N}) \triangleq \Prob{S_0 = s_0, S_N = s_N, S_{2N} = s_{2N}} .
	\end{equation}
\begin{lemma}
	\label{lemm:positive probability of a triplet}
	There exists $N_0$ and $\mu > 0$ such that for all $N \geq N_0$ and all $(s_0, s_N, s_{2N}) \in \mathcal{S}_N^3$ we have
	\[
		\pi(s_0, s_N, s_{2N}) > \mu .
	\]
\end{lemma}
\begin{IEEEproof} See Appendix~\ref{app:auxiliary proofs}. \end{IEEEproof}

For a given $0 \leq \Delta < \deriod$,  $N=2^n \geq \deriod$, and  $1 \leq i \leq N$, we follow \cite[sec. VI.B]{TPFV:22p} and define for $(s_0,s_N,s_{2N}) \in \mathcal{S}_N^3(\Delta)$,
\begin{multline}
	\label{eq:alpha}
	\alpha(s_0,s_N,s_{2N}) \triangleq \\
H(U_i + V_i | Q_i, R_i, S_0 = s_0, S_N = s_N, S_{2N} = s_{2N} ) 
\end{multline}
and
\begin{equation}
	\label{eq:beta}
	\beta(s_0,s_N,s_{2N}) \triangleq \frac{\gamma(s_0,s_N) + \gamma(s_N,s_{2N})}{2} ,
\end{equation}
where
\begin{equation}
	\label{eq:gamma}
\gamma(s_0,s_N) \triangleq H(U_i|Q_i, S_0 = s_0, S_N = s_N) .
\end{equation}
By stationarity,
\[
\gamma(s_N,s_{2N}) = H(V_i|R_i, S_N = s_N, S_{2N} = s_{2N}) .
\]
The following is the analog of \cite[Lemma 16]{TPFV:22p} for our setting.
\begin{lemma}
	\label{lemm:HnHnMinusInTermsOfAlphaBeta}
	Let $N = 2^n \geq \max\{N_0,\deriod\}$, for $N_0$ as in \Cref{lemm:positive probability of a triplet}. For $0 \leq \Delta < \deriod$ and $1 \leq i \leq N$ fixed
	\begin{equation}
		\label{eq:Hnminus alpha}
		\hat{H}_n^-(\Delta) \geq \deriod \cdot\sum_{\mathclap{(s_0,s_N,s_{2N}) \in \mathcal{S}_N^3(\Delta)}} \pi(s_0,s_N,s_{2N}) \cdot \alpha(s_0,s_N,s_{2N}) ,
	\end{equation}	
	and
	\begin{equation}
		\label{eq:Hn beta}
		\hat{H}_n(\Delta) = \deriod \cdot\sum_{\mathclap{(s_0,s_N,s_{2N}) \in \mathcal{S}_N^3(\Delta)}} \pi(s_0,s_N,s_{2N}) \cdot \beta(s_0,s_N,s_{2N}) .
	\end{equation}	
	Furthermore, for all $(s_0,s_N,s_{2N}) \in \mathcal{S}_N^3(\Delta)$,
	\begin{equation}
		\label{eq:alpha beta}
\alpha(s_0,s_N,s_{2N}) \geq \beta(s_0,s_N,s_{2N}) .
	\end{equation}
\end{lemma}

\begin{IEEEproof}
	Note that under the stationary distribution, we have
	\[ 
		\Prob{ S_0 = s_0, S_N = s_N, S_{2N} = s_{2N} | D(\Delta) } = \deriod \cdot \pi(s_0,s_N,s_{2N}). 
	\] 
	Hence, by \eqref{eq:HnMinusDelta}, \eqref{eq:alpha}, and since conditioning reduces entropy,
	\begin{IEEEeqnarray*}{rCl}
		\hat{H}_n^-(\Delta) & = & H(U_i+V_i|Q_i,R_i, S_0, S_{2N}, D(\Delta)) \\
			      & \geq & H(U_i+V_i|Q_i,R_i, S_0, S_N, S_{2N}, D(\Delta)) \\
			      & = & \sum_{\mathclap{(s_0,s_N,s_{2N}) \in \mathcal{S}_N^3(\Delta)}} \deriod \cdot \pi(s_0,s_N,s_{2N}) \cdot \alpha(s_0,s_N,s_{2N}), 
\end{IEEEeqnarray*}
	which establishes~\eqref{eq:Hnminus alpha}. 

	To see~\eqref{eq:Hn beta}, observe that by stationarity we have
	\begin{IEEEeqnarray*}{rCl}
		\hat{H}_n(\Delta) & = & H(U_i|Q_i, S_0, S_N, D(\Delta)) \\ 
				  & = & \frac{H(U_i|Q_i, S_0, S_N, D(\Delta)) +H(V_i|R_i, S_N, S_{2N}, D(\Delta))}{2} \\ 
		    & = & \sum_{\mathclap{(s_0,s_N,s_{2N}) \in \mathcal{S}_N^3(\Delta)}} \deriod \cdot \pi(s_0,s_N,s_{2N}) \cdot \frac{\gamma(s_0,s_N)+\gamma(s_N,s_{2N})}{2} \\ 
		    & = & \sum_{\mathclap{(s_0,s_N,s_{2N}) \in \mathcal{S}_N^3(\Delta)}} \deriod \cdot \pi(s_0,s_N,s_{2N}) \cdot \frac{\beta(s_0,s_N,s_{2N})}{2},
\end{IEEEeqnarray*}
	where in the last two equalities we used~\eqref{eq:gamma} and~\eqref{eq:beta}.
	
	Finally, 
	\begin{IEEEeqnarray*}{rCl}
		\IEEEeqnarraymulticol{3}{l}{\alpha(s_0,s_N,s_{2N})} \\ 
		&\eqann[\geq]{a}& H(U_i+V_i|V_i, Q_i,R_i, S_0 = s_0, S_N = s_N, S_{2N} = s_{2N}) \\ 
		&=& H(U_i|V_i, Q_i,R_i, S_0 = s_0, S_N = s_N, S_{2N} = s_{2N}) \\ 
		&\eqann{b}& H(U_i|Q_i, S_0 = s_0, S_N = s_N) \\ 
		& = & \gamma(s_0,s_N), 
\end{IEEEeqnarray*}
	where \eqannref{a} is since conditioning by $V_i$ reduces entropy, and \eqannref{b} is by the Markov property. Similarly, by conditioning on $U_i$, we have 
	\[ 
\alpha(s_0,s_N,s_{2N}) \!\geq\! H(V_i|R_i, S_N = s_N, S_{2N} = s_{2N}) \!=\! \gamma(s_N,s_{2N}).
\] 
Combining these two inequalities and recalling~\eqref{eq:beta}  yields~\eqref{eq:alpha beta}, which completes the proof.
\end{IEEEproof}
	
Recall from~\eqref{eq:period factorization} the factorization $\period = \deriod \cdot \qeriod$, where $\deriod$ is a power of $2$ and $\qeriod$ is odd. We have already used the special structure of $\deriod$ previously. The \emph{entire} factorization of $\period$ is the key for proving the periodic analog of \cite[Lemma 17]{TPFV:22p}, below.

\begin{lemma}
\label{lemm:alpha strictly greater than beta}
For every $0 < \epsilon < 1$ there exists $\Theta= \Theta(\epsilon) > 0$, for which the following holds. Let $N = 2^n \geq \max\{N_0,\deriod\}$, where the Euler totient of $\qeriod$ divides $n$, and $N_0$ is as promised in \Cref{lemm:positive probability of a triplet}. Fix $0 \leq \Delta  < \deriod$ and $1 \leq i \leq N$. Then, if $\epsilon \leq \hat{H}_n(\Delta) \leq 1 - \epsilon$, there exist $(s_0,s_N, s_{2N}) \in \mathcal{S}_N^3(\Delta)$ such that
	\begin{equation}
		\label{eq:alpha strictly greater than beta}
		\alpha(s_0,s_N,s_{2N}) > \beta(s_0,s_N,s_{2N}) + \Theta .
	\end{equation}
\end{lemma}

\begin{IEEEproof}
	By definition of $\gamma$ in \eqref{eq:gamma} and stationarity,
	\begin{IEEEeqnarray}{rCl}
		\hat{H}_n(\Delta) &=& \deriod \sum_{\mathclap{(s_0,s_N) \in \mathcal{S}_N^2(\Delta)}} \Prob{S_0 = s_0, S_N = s_N} \cdot \gamma(s_0, s_N) \label{eq:HnDelta as sum of gamma} \\
			    &=& \deriod \sum_{\mathclap{(s_N,s_{2N}) \in \mathcal{S}_N^2(\Delta)}} \Prob{S_N = s_N, S_{2N} = s_{2N}} \cdot \gamma(s_N,s_{2N}) \IEEEnonumber .
	\end{IEEEeqnarray}

	Our aim is to show that there exists a triplet $(s_0,s_N,s_{2N}) \in \mathcal{S}_N^3(\Delta)$ for which
	\begin{equation}
		\label{eq:min of gamma pair}
		\min\{\gamma(s_0,s_N), \gamma(s_N, s_{2N})\} \leq 1 - \epsilon
	\end{equation}
	and
	\begin{equation}
		\label{eq:max of gamma pair}
		\max\{\gamma(s_0,s_N), \gamma(s_N, s_{2N})\} \geq \epsilon .
	\end{equation}
The above would imply \eqref{eq:alpha strictly greater than beta}, due to \cite[Lemma 11]{SasogluTal:19p} and part (i) of \cite[Lemma 2.2]{Sasoglu:12b}. Note that the latter is required since the former does not involve conditioning. We further note that \cite[Lemma 11]{SasogluTal:19p} is stated for strict inequalities, but actually holds also for non-strict inequalities, as can be seen in its proof.

We will show the existence of a triplet $(s_0,s_N,s_{2N}) \in \mathcal{S}_N^3(\Delta)$ for which
\begin{equation}
	\gamma(s_0,s_N) \leq \hat{H}_n(\Delta) \quad \mbox{and} \quad \gamma(s_N, s_{2N}) \geq \hat{H}_n(\Delta)
\end{equation}
or
\begin{equation}
	\gamma(s_0,s_N) \geq \hat{H}_n(\Delta) \quad \mbox{and} \quad \gamma(s_N, s_{2N}) \leq \hat{H}_n(\Delta)
\end{equation}
Note that by our assumption that $\epsilon \leq \hat{H}_n(\Delta) \leq 1-\epsilon$, the above would indeed imply \eqref{eq:min of gamma pair} and \eqref{eq:max of gamma pair}. The above can be written succinctly as
\[
	(\hat{H}_n(\Delta) - \gamma(s_0,s_N)) \cdot (\hat{H}_n(\Delta) - \gamma(s_N,s_{2N})) \leq 0 .
\]

The proof is by contradiction. Namely, assume that for all $(s_0,s_N,s_{2N}) \in \mathcal{S}_N^3(\Delta)$,
\begin{equation}
	\label{eq:Hn contradiction}
	(\hat{H}_n(\Delta) - \gamma(s_0,s_N)) \cdot (\hat{H}_n(\Delta) - \gamma(s_N,s_{2N})) > 0 .
\end{equation}

Let phase $0 \leq \varphi < \period = \deriod \cdot \qeriod$ be such that $\varphi \equiv \Delta \pmod{\deriod}$ and $\varphi \equiv -2 \pmod{\qeriod}$. Such a $\varphi$ indeed exists, by the Chinese remainder theorem \cite[Theorem 2.12.6]{PaleyWeichsel:66b}. Fix $s_0,s_N$ such that $\phi(s_0) = \varphi$ and $\phi(s_N) \equiv \varphi + N \pmod{\period}$. Note that $(s_0,s_N) \in \mathcal{S}_N^2(\Delta)$. For $s \in \mathcal{S}$, denote $\phi'(s) = \phi(s) \bmod \qeriod$. Thus, $\phi'(s_0) \equiv -2 \pmod{\qeriod}$ and $\phi'(s_N) \equiv -1 \pmod{\qeriod}$, where the second equality is by Euler's theorem \cite[Theorem 2.14.1]{PaleyWeichsel:66b} and our assumption that the totient of $\qeriod$ divides $n$.
Since \eqref{eq:Hn contradiction} is assumed to hold, w.l.o.g.\ further assume that $\gamma(s_0,s_N) < \hat{H}_n(\Delta)$. 

Our aim is to show that for any $(s,s') \in \mathcal{S}_N^2(\Delta)$, we have $\gamma(s,s') < \hat{H}_n(\Delta)$. This would establish the contradiction, by \eqref{eq:HnDelta as sum of gamma}. Indeed, by stationarity, for every $0 \leq \Delta < \deriod$, $\sum_{(s,s') \in \mathcal{S}_N^2(\Delta)} \Prob{S_0 = s, S_N = s'} = 1/\deriod$. We do this by showing that for any $0 \leq j < \qeriod$ and any $(s,s') \in \mathcal{S}_N^2(\Delta)$ such that $\phi'(s) = j$, we have $\gamma(s,s') < \hat{H}_n(\Delta)$.
\begin{enumerate}
	\item We have already established that $\gamma(s_0,s_N) < \hat{H}_n(\Delta)$. Recall that $\phi'(s_0) = -2$ and $\phi'(s_N) \equiv -1 \pmod{\qeriod}$.
	\item \label{it:sNa2} We claim that for any $a_0 \in \mathcal{S}$ such that $(s_N, a_0) \in \mathcal{S}_N^2(\Delta)$, we have $\gamma(s_N, a_0) < \hat{H}_n(\Delta)$. Indeed, this follows by \eqref{eq:Hn contradiction}, and noting that $(s_0,s_N,a_0) \in \mathcal{S}_N^3(\Delta)$. Note that $\phi'(a_0) \equiv 0 \pmod{\qeriod}$.
	\item \label{it:sNa2a3} We now claim that for any $a_0, a_1$  such that $(s_N,a_0,a_1) \in \mathcal{S}_N^3(\Delta)$, we have $\gamma(a_0,a_1) < \hat{H}_n(\Delta)$. Indeed, this follows by \eqref{eq:Hn contradiction} and item \ref{it:sNa2}. Note that $\phi'(a_0) \equiv 0 \pmod{\qeriod}$ and $\phi'(a_1) \equiv 1 \pmod{\qeriod}$.
	\item We conclude from item \ref{it:sNa2a3} that for any $(a_0,a_1) \in \mathcal{S}_N^2(\Delta)$ such that $\phi'(a_0) \equiv 0 \pmod{\qeriod}$ we have $\gamma(a_0,a_1) < \hat{H}_n(\Delta)$. This will be the basis of our induction on $j$. Note that $\phi'(a_1) \equiv 1 \pmod{\qeriod}$.
	\item For the induction step, assume that for some $j \geq 0$, for any $(a_j,a_{j+1}) \in \mathcal{S}_N^2(\Delta)$ such that $\phi'(a_j) \equiv j \pmod{\qeriod}$, it holds that $\gamma(a_j, a_{j+1}) < \hat{H}_n(\Delta)$. We now show that for any $(a_{j+1},a_{j+2}) \in \mathcal{S}_N^2(\Delta)$ such that $\phi'(a_{j+1}) \equiv j + 1 \pmod{\qeriod}$, it holds that $\gamma(a_{j+1},a_{j+2}) < \hat{H}_n(\Delta)$. Indeed this holds since $(a_j,a_{j+1},a_{j+2}) \in \mathcal{S}_N^3(\Delta)$, the induction hypothesis, and~\eqref{eq:Hn contradiction}. 
\end{enumerate}
This completes the proof.
\end{IEEEproof}

The following is a corollary of \Cref{lemm:positive probability of a triplet,lemm:HnHnMinusInTermsOfAlphaBeta,lemm:alpha strictly greater than beta}.

\begin{corollary}
	\label{coro:HnMinusIncreases}
	Fix $0<\epsilon < 1$. There exists $\tau(\epsilon) > 0$ such that the following holds. If $N = 2^n$ is as in \Cref{lemm:alpha strictly greater than beta}, $0 \leq \Delta  < \deriod$, $1 \leq i \leq N$, and $\epsilon \leq \hat{H}_n(\Delta) \leq 1 - \epsilon$, then
	\[
		\hat{H}_n^-(\Delta) - \hat{H}_n(\Delta) > \tau(\epsilon) .
	\]
\end{corollary}
\begin{IEEEproof}
	Take $\tau(\epsilon) = \mu \cdot \deriod \cdot \Theta(\epsilon)$, where $\mu$ is as promised in \Cref{lemm:positive probability of a triplet} and $\Theta(\epsilon)$ is as promised in \Cref{lemm:alpha strictly greater than beta}.
\end{IEEEproof}

The following lemma is essentially \cite[Theorem 1]{Sasoglu:11z}, adapted to our case.
\begin{lemma}
	\label{lemm:DDelta converges to zero one}
	The random variable $\hat{H}_\infty(\Delta)$ is a $\{0,1\}$ random variable with probability $1$.
\end{lemma}

\begin{IEEEproof}
	Consider the random process $(\hat{H}_n(\Delta))_{n=1}^\infty$, as in \Cref{lemm:H_n converges}, where it was shown to converge a.s., in $L_1$, and in probability to $\hat{H}_\infty(\Delta) \in [0,1]$. To show that $\hat{H}_\infty(\Delta) \in \{0,1\}$ with probability $1$,  assume to the contrary that there exists $\epsilon > 0$ such $\Prob{\hat{H}_\infty(\Delta) \in (2\epsilon,1-2\epsilon)} > \rho$, for some $\rho > 0$. Next, note that
	\begin{IEEEeqnarray*}{rCl}
		\IEEEeqnarraymulticol{3}{l}{\Prob{ \hat{H}_n(\Delta) \in (\epsilon,1-\epsilon) }} \\
		\quad &\geq& \Prob{\hat{H}_n(\Delta) \in (\epsilon,1- \epsilon) \quad \mbox{and} \quad |\hat{H}_n(\Delta) - \hat{H}_\infty(\Delta)| < \epsilon} \\
		      &\geq& \Prob{\hat{H}_\infty(\Delta) \in (2\epsilon,1-2\epsilon) \quad \mbox{and} \quad |\hat{H}_n(\Delta) - H_\infty(\Delta)| < \epsilon} \\
		      &=& \Prob{\hat{H}_\infty(\Delta) \in (2\epsilon, 1- 2 \epsilon)} \\
		      && - \Prob{\hat{H}_\infty(\Delta) \in (2\epsilon, 1 - 2\epsilon) \quad \mbox{and} \quad |\hat{H}_n(\Delta) - H_\infty(\Delta)| \geq \epsilon} \\
		      &\geq& \Prob{\hat{H}_\infty(\Delta) \in (2\epsilon, 1- 2 \epsilon)} - \Prob{ |\hat{H}_n(\Delta) - H_\infty(\Delta)| \geq \epsilon} \\
		      & > & \rho  - \Prob{ |\hat{H}_n(\Delta) - \hat{H}_\infty(\Delta)| \geq \epsilon} ,
	\end{IEEEeqnarray*}
	where the last inequality follows from our contradictory assumption. By convergence in probability of $\hat{H}_n(\Delta)$, we obtain
	\[
		\liminf_{n \to \infty} \Prob{\hat{H}_n(\Delta) \in (\epsilon,1-\epsilon)} \geq \rho .
	\]
	Thus, from \Cref{coro:HnMinusIncreases}, recalling our constraint on $n$ from \Cref{lemm:alpha strictly greater than beta}, there exist an infinite sequence $n = n_1, n_2, \ldots$ and $\tau > 0$ such that $\Prob{|\hat{H}_n^-(\Delta) - \hat{H}_n(\Delta)| > \tau } > \rho/2$. Since $\hat{H}_{n+1}(\Delta)$ equals $\hat{H}_n^-(\Delta)$ with probability $1/2$, we have $\Prob{|\hat{H}_{n+1}(\Delta) - \hat{H}_n(\Delta)| > \tau } > \rho/4$. This implies that $\hat{H}_n(\Delta)$ cannot converge in probability. This is a contradiction, and we conclude that $\hat{H}_\infty(\Delta) \in \{0,1\}$ with probability $1$.
\end{IEEEproof}

The following lemma states that fixing the initial phase does not change the entropy rate. Note that in the following we have already established the existence of the limit on the left-hand side, and claim that the right-hand side limit exists.
\begin{lemma} \label{lemm:entropy rate with and without initial phase}
	For $0 \leq \varphi < \period$, we have
	\[
		\lim_{N \to \infty} \frac{1}{N} H(X_1^N| Y_1^N) = \lim_{N \to \infty} \frac{1}{N} H(X_1^N| Y_1^N, \phi(S_0) = \varphi) . 
	\]
\end{lemma}

\begin{IEEEproof}
	By \eqref{eq:technical simeq relations}, we have $H(X_1^N|Y_1^N) \simeq H(X_1^N|Y_1^N, \phi(S_0))$, by taking $A$ as the always true event. Thus, our aim is to show that for any $0 \leq \varphi < \period$,
	\[
H(X_1^N|Y_1^N, \phi(S_0)) \simeq H(X_1^N| Y_1^N, \phi(S_0) = \varphi).
	\]
	By stationarity, we have $\Prob{\phi(S_0) = -\varphi' \bmod{\period}} = 1/\period$ for any $0 \leq \varphi' < \period$. Thus,
		\begin{equation}
			\label{eq:Hphi as sum}
			H(X_1^N| Y_1^N, \phi(S_0)) = \sum_{\varphi' = 0}^{\period -1} \frac{1}{\period} H(X_1^N| Y_1^N, \phi(S_0) = -\varphi' \bmod{\period}) .
	\end{equation}
	Recall that $\varphi$ is fixed. For a given $0 \leq \varphi'\leq \period$, denote $\psi = \varphi' + \varphi$. Note that $\psi \geq 0$. We have
\begin{IEEEeqnarray*}{rCl}
	\IEEEeqnarraymulticol{3}{l}{H(X_1^N|Y_1^N, \phi(S_0) \equiv -\varphi' \bmod{\period})} \\
	\quad &\eqann[\simeq]{a}& H(X_1^N|Y_1^N, S_\psi , S_N, \phi(S_0) \equiv -\varphi' \bmod{\period}) \\
      &=& H(X_1^N|Y_1^N, S_\psi , S_N, \phi(S_\psi) = \varphi) \\
      & \eqann{b} & H(X_1^N|Y_1^{N+\psi}, S_\psi , S_N, \phi(S_\psi) = \varphi) \\
      & \eqann[\simeq]{c} & H(X_1^{N+\psi}|Y_1^{N+\psi}, S_\psi , S_N, \phi(S_\psi) = \varphi) \\
      & \eqann[\simeq]{d} & H(X_{\psi + 1}^{N+\psi}|Y_{1}^{N+\psi}, S_\psi , S_N, \phi(S_\psi) = \varphi) \\
      & \eqann{e} & H(X_{\psi + 1}^{N+\psi}|Y_{\psi + 1}^{N+\psi}, S_\psi , S_N, \phi(S_\psi) = \varphi) \\
      & \eqann[\simeq]{f}  & H(X_{\psi + 1}^{N+\psi}|Y_{\psi + 1}^{N+\psi},  \phi(S_\psi) = \varphi) \\
      & \eqann{g} & H(X_{1}^{N}|Y_{1}^{N}, \phi(S_0) = \varphi),
\end{IEEEeqnarray*}
where \eqannref{a} is by the RHS of~\eqref{eq:technical simeq relations} since the number of states is finite; \eqannref{b} is by the Markov property; \eqannref{c} is by the LHS of~\eqref{eq:technical simeq relations} since we have added a finite number $\psi$ of  binary random variables;  \eqannref{d} is again by the LHS of~\eqref{eq:technical simeq relations}, now removing a finite number $\psi$ of binary random variables; \eqannref{e} is by the Markov property; \eqannref{f} is by the RHS of~\eqref{eq:technical simeq relations}; and \eqannref{g} is by stationarity.

Thus, we have shown that
\[
	H(X_1^N|Y_1^N, \phi(S_0) = -\varphi' \bmod{\period}) \simeq H(X_{1}^{N}|Y_{1}^{N}, \phi(S_0) = \varphi).
\]
Using this in~\eqref{eq:Hphi as sum} we obtain 
\[
	H(X_1^N|Y_1^N, \phi(S_0)) \simeq H(X_{1}^{N}|Y_{1}^{N}, \phi(S_0) = \varphi), 
\] 
completing the proof.
\end{IEEEproof}

The following proposition establishes slow polarization. It is in a similar vein to \cite[Theorem 20]{TPFV:22p}.
\begin{proposition} \label{prop:slow polarization}
For $0 \leq \Delta < \deriod$, $0<\epsilon < 1$, and $N=2^n$, the following holds:
\begin{IEEEeqnarray}{rCl} 
	\IEEEeqnarraymulticol{3}{l}{\lim_{n \to \infty} \frac{1}{N} \left|\left\{ i : H(U_i | U_1^{i-1}, Y_1^N, D(\Delta) ) < \epsilon \right\}\right|}
		\IEEEyesnumber \label{eq:one minus entropy rate chain} \IEEEyessubnumber \label{eqs:limit set D epsilon} \\
		&= & \lim_{n \to \infty} \frac{1}{N} \left|\left\{ i : H(U_i | U_1^{i-1}, Y_1^N, S_0, S_N,  D(\Delta) ) < \epsilon \right\}\right|\IEEEyessubnumber \label{eqs:limit set S0 SN D epsilon} \IEEEeqnarraynumspace \\
		&= & 1 - \lim_{n \to \infty} \frac{1}{N} H(X_1^N|Y_1^N, S_0, S_N, D(\Delta)) \IEEEyessubnumber \label{eqs:one minus entropy rate S0 SN D} \\
		&= & 1 - \lim_{n \to \infty} \frac{1}{N} H(X_1^N|Y_1^N, D(\Delta)) \IEEEyessubnumber \label{eqs:one minus entropy rate D} \\
		&= & 1 - \lim_{n \to \infty} \frac{1}{N} H(X_1^N|Y_1^N) \IEEEyessubnumber \label{eqs:one minus entropy rate} 
	\end{IEEEeqnarray}
	and
	\begin{IEEEeqnarray}{rCl}
		\IEEEeqnarraymulticol{3}{l}{\lim_{n \to \infty} \frac{1}{N} \left|\left\{ i : H(U_i | U_1^{i-1}, Y_1^N, D(\Delta) ) > 1 -  \epsilon \right\}\right|} 
		\IEEEyesnumber \label{eq:entropy rate chain} \IEEEyessubnumber \label{eqs:limit set D one minus epsilon}
		\\
		&= & \lim_{n \to \infty} \frac{1}{N} \left|\left\{ i : H(U_i | U_1^{i-1}, Y_1^N, S_0, S_N,  D(\Delta) ) > 1 - \epsilon \right\}\right| \IEEEyessubnumber \label{eqs:limit set S0 SN D one minus epsilon}\IEEEeqnarraynumspace\\
		&= & \lim_{n \to \infty} \frac{1}{N} H(X_1^N|Y_1^N, S_0, S_N, D(\Delta))\IEEEyessubnumber \label{eqs:entropy rate S0 SN D} \\
		&= & \lim_{n \to \infty} \frac{1}{N} H(X_1^N|Y_1^N, D(\Delta)) \IEEEyessubnumber \label{eqs:entropy rate D}\\
		&= & \lim_{n \to \infty} \frac{1}{N} H(X_1^N|Y_1^N) . \IEEEyessubnumber \label{eqs:entropy rate} 
	\end{IEEEeqnarray}
\end{proposition}

\begin{IEEEproof}
We first prove \eqref{eqs:entropy rate}=\eqref{eqs:entropy rate D}, which implies \eqref{eqs:one minus entropy rate}=\eqref{eqs:one minus entropy rate D}. Note that $D(\Delta)$ is a union of the events $\phi(S_0) = \varphi$, where $\varphi \equiv \Delta \pmod \deriod$. Given $D(\Delta)$, each such event has probability $1/\qeriod$, by stationarity. Hence,
\[
	H(X_1^N|Y_1^N,D(\Delta)) = \sum_{\varphi \equiv \Delta\!\!\!\!\! \pmod \deriod} \frac{1}{\qeriod} H(X_1^N|Y_1^N,\phi(S_0) = \varphi).
\]
Note that the number of summands is $\qeriod$, by the Chinese remainder theorem \cite[Theorem 2.12.6]{PaleyWeichsel:66b}. Multiplying both sides by $1/N$ and taking limits with $n \to \infty$, we use \Cref{lemm:entropy rate with and without initial phase} to obtain the desired equality.

We next prove \eqref{eqs:entropy rate D}=\eqref{eqs:entropy rate S0 SN D}, which implies \eqref{eqs:one minus entropy rate D}=\eqref{eqs:one minus entropy rate S0 SN D}. We do so by using the notation $\simeq$ from~\Cref{def:simeq}. By the RHS of
\eqref{eq:technical simeq relations}, and noting that the number of states is finite, we have $H(X_1^N|Y_1^N, D(\Delta)) \simeq H(X_1^N|Y_1^N, S_0, S_N, D(\Delta))$, from which the equality follows.

We next prove that \eqref{eqs:entropy rate S0 SN D}=\eqref{eqs:limit set S0 SN D one minus epsilon} and \eqref{eqs:one minus entropy rate S0 SN D}=\eqref{eqs:limit set S0 SN D epsilon}. First, note that for the random process $(\hat{H}_n(\Delta))_{n=1}^\infty$, as in \Cref{lemm:H_n converges},
\begin{IEEEeqnarray*}{rCl}
	\Exp{\hat{H}_n(\Delta)} &=& \frac{1}{N} \sum_{i=1}^N H(U_i | U_1^{i-1}, Y_1^N, S_0, S_N,  D(\Delta) ) \\
			  &=& \frac{1}{N} H(U_1^N | Y_1^N, S_0, S_N,  D(\Delta) ) \\
			  &=& \frac{1}{N} H(X_1^N | Y_1^N, S_0, S_N,  D(\Delta) ) .
\end{IEEEeqnarray*}
Thus, $\lim_{n \to \infty} \Exp{\hat{H}_n(\Delta)}$ equals \eqref{eqs:entropy rate S0 SN D}. By the bounded convergence theorem \cite[Theorem 7.3.1]{Rosenthal:06b} and
\Cref{lemm:H_n converges}, 
\begin{multline} \label{eq:Exp Hinfty Delta}\lim_{n \to \infty} \Exp{\hat{H}_n(\Delta)} =  \Exp{\lim_{n \to \infty} \hat{H}_n(\Delta)} = \Exp{\hat{H}_\infty(\Delta)} \\ = \Prob{\hat{H}_\infty(\Delta) = 1} = \Prob{\hat{H}_{\infty}(\Delta) > 1- \epsilon},\end{multline} 
where the latter two equalities are since $\hat{H}_\infty(\Delta) \in \{0,1\}$ by \Cref{lemm:HnHnMinusInTermsOfAlphaBeta}. Hence, 
\begin{equation} \label{eq:displayed equation for 23c=23d} 
	\begin{IEEEeqnarraybox}[][c]{rCl}
		\eqref{eqs:entropy rate S0 SN D} &=& \Prob{\hat{H}_{\infty}(\Delta) > 1-\epsilon}  \\  & \eqann{a}& \lim_{n\to\infty} \Prob{\hat{H}_n(\Delta) > 1-\epsilon} = \eqref{eqs:limit set S0 SN D one minus epsilon} , 
\end{IEEEeqnarraybox}
\end{equation} 
where~\eqannref{a} is by \cite[Theorem 10.1.1(3)]{Rosenthal:06b} and since convergence in probability of $\hat{H}_n(\Delta)$ implies convergence in distribution \cite[Theorem 10.2.1]{Rosenthal:06b}. This establishes \eqref{eqs:limit set S0 SN D one minus epsilon}=\eqref{eqs:entropy rate S0 SN D}. The equality 
\eqref{eqs:limit set S0 SN D epsilon} = \eqref{eqs:one minus entropy rate S0 SN D} follows similarly, for the process $(1-\hat{H}_n(\Delta))_n$. 

Finally, we prove that \eqref{eqs:entropy rate S0 SN D}=\eqref{eqs:limit set D one minus epsilon} and \eqref{eqs:one minus entropy rate S0 SN D}=\eqref{eqs:limit set D epsilon}. To this end, we define the random sequence $(H_n(\Delta))_{n=1}^{\infty}$, where $H_n(\Delta)$ was defined in~\eqref{eq:HnBarDelta} and the index $i$ is drawn in the same manner as in \Cref{lemm:H_n converges}. Now, utilizing the chain rule we obtain that 
\begin{multline*} \Exp{H_n(\Delta)} = \frac{1}{N}\sum_{i=1}^N H(U_i|U_1^{i-1},Y_1^N,D(\Delta)) \\= \frac{1}{N}H(U_1^N|Y_1^N,D(\Delta)) = \frac{1}{N}H(X_1^N|Y_1^N,D(\Delta)).  \end{multline*}
Taking limits on both sides yields that 
\[
	\lim_{n\to\infty} \Exp{H_n(\Delta)} = \lim_{n\to\infty} \frac{1}{N} H(X_1^N | Y_1^N, D(\Delta)) = \Exp{\hat{H}_{\infty}(\Delta)}, 
\]
where the latter equality is since we have shown above that \eqref{eqs:entropy rate D} $=$ \eqref{eqs:entropy rate S0 SN D} $=$  \eqref{eqs:limit set S0 SN D one minus epsilon} $= \Exp{\hat{H}_{\infty}(\Delta)}$, see~\eqref{eq:Exp Hinfty Delta} and \eqref{eq:displayed equation for 23c=23d}.
Next, since conditioning reduces entropy we have $H_n(\Delta) \geq \hat{H}_n(\Delta)$. By \cite[Lemma 19]{TPFV:22p}, this implies that $H_n(\Delta)$ converges in $L_1$ to $\hat{H}_{\infty}(\Delta)$. Since $H_n(\Delta)$ is also bounded, then $L_1$ convergence implies convergence in probability and thus in distribution. Now, using the same argument as for~\eqref{eq:displayed equation for 23c=23d}, 
\[ 
	\eqref{eqs:entropy rate S0 SN D} = \Prob{\hat{H}_{\infty}(\Delta) > 1-\epsilon} = \lim_{n\to \infty}\Prob{H_n(\Delta) > 1-\epsilon} = \eqref{eqs:limit set D one minus epsilon}. 
\]
To see that \eqref{eqs:one minus entropy rate S0 SN D} $=$ \eqref{eqs:limit set D epsilon}, we repeat the above for the process $(1-H_n(\Delta))_n$. 
\end{IEEEproof}

\section{Proof of Fast Polarization}
To prove fast polarization, we first define $\mixing(\Delta)$ for all $0 \leq \Delta < \deriod$,
	\begin{equation} \label{eq:mixing def}
		\mixing(\Delta) = \frac{1}{\deriod \cdot \min\{\pi_s : s \in \mathcal{S}(\Delta)\}}.  
\end{equation}
Recall that for any $s$, we have by irreducibility and stationarity that $\pi_s >0$, and hence $\mixing(\Delta)$ is well defined (the denominator is positive). The utility of $\mixing(\Delta)$ is seen in the next lemma, which allows us to bound the probability of two dependent events by their marginal probabilities.

\begin{lemma}\label{lemm:mixing}
	Fix $0 \leq \Delta < \deriod$ and $N=2^n \geq \deriod$. Let $A, B \subseteq \mathcal{X}^N \times \mathcal{Y}^N$. Then,
	\begin{multline}
		\label{eq:mixingAB}
		\Prob{(X_1^N,Y_1^N) \in A \quad \mbox{and} \quad (X_{N+1}^{2N}, Y_{N+1}^{2N}) \in B \Big| D(\Delta) } \\
		\leq \mixing(\Delta) \cdot \Prob{(X_1^N,Y_1^N) \in A \Big| D(\Delta) )} \\
		\cdot \Prob{(X_{N+1}^{2N}, Y_{N+1}^{2N}) \in B \Big| D(\Delta) } .
	\end{multline}
\end{lemma}

\begin{IEEEproof}
	See Appendix~\ref{app:auxiliary proofs}.
\end{IEEEproof}

Recall the definitions of $U_i,V_i,Q_i,R_i$ from \Cref{sec:setting}. 
For a given $0 \leq \Delta  < \deriod$, $N = 2 ^n$, and $1 \leq i \leq N$, denote
\begin{IEEEeqnarray*}{rCl}
	Z_n(\Delta) &=& Z(U_i|Q_i,D(\Delta)) \\
	Z_n^-(\Delta) &=& Z(U_i + V_i|Q_i, R_i, D(\Delta)) \\
	Z_n^+(\Delta) &=& Z(V_i| U_i + V_i,Q_i,R_i, D(\Delta)) 
\end{IEEEeqnarray*}
and
\begin{IEEEeqnarray*}{rCl}
	\hat{K}_n(\Delta) &=& K(U_i|Q_i, S_0, S_N, D(\Delta)) \\
	\hat{K}_n^-(\Delta) &=& K(U_i + V_i|Q_i,R_i, S_0, S_{2N}, D(\Delta)) \\
	\hat{K}_n^+(\Delta) &=& K(V_i| U_i + V_i,Q_i,R_i, S_0, S_{2N}, D(\Delta)).
\end{IEEEeqnarray*}

\begin{lemma} \label{lemm:Zn satisfies 167}
Fix $0 \leq \Delta  < \deriod$, $N = 2 ^n \geq \deriod$, and $1 \leq i \leq N$. Then,
\[
	Z_n^-(\Delta) \leq 2 \cdot \mixing(\Delta) \cdot Z_n(\Delta) \quad \mbox{and} \quad Z_n^+(\Delta) \leq \mixing(\Delta) \cdot (Z_n(\Delta))^2 .
\]
\end{lemma}
This is an extension of~\cite[Section VI]{SasogluTal:19p} for our scenario.
\begin{IEEEproof}
	We first prove that $Z_n^-(\Delta) \leq 2 \cdot \mixing(\Delta) \cdot Z_n(\Delta)$. As a shorthand, denote for an $i$ clear from the context,
	\begin{IEEEeqnarray*}{rCl}
		p( u , q ) &=& \Prob{U_i = u, Q_i = q| D(\Delta) } \\
		p( v , r ) &=& \Prob{V_i = v, R_i = r| D(\Delta) } \\
		p( u, v , q, r ) &=& \Prob{U_i = u, V_i = v, Q_i = q,  R_i = r| D(\Delta) }.
	\end{IEEEeqnarray*}
	We further denote 
	\begin{equation}\label{eq:abcd def}
		\begin{IEEEeqnarraybox}[][c]{rClrCl}
			a_q &=& p(0,q), &\quad  b_q &=& p(1,q), \\
			c_r &=& p(0,r), &\quad d_r &=& p(1,r).
		\end{IEEEeqnarraybox}
	\end{equation}
	Observe that by \Cref{lemm:mixing}, we have 
	\begin{equation} \label{eq:puvqr MDelta}
	p(u,v,q,r) \leq M(\Delta) p(u,q) p(v,r). 
\end{equation}

	Indeed,
	\begin{IEEEeqnarray*}{rCl}
		Z_n^-(\Delta) & = & 2\sum_{q,r} \sqrt{\Prob{U_i + V_i = 0, Q_i = q, R_i = r| D(\Delta)}} \\
		\IEEEeqnarraymulticol{3}{r}{\cdot \sqrt{\Prob{U_i + V_i = 1, Q_i = q, R_i = r| D(\Delta)}}} \\
			      &=& 2 \sum_{q,r} \sqrt{\left(p(0,0,q,r) + p(1,1,q,r)\right)} \\
		\IEEEeqnarraymulticol{3}{r}{\cdot \sqrt{\left(p(0,1,q,r) + p(1,0,q,r)\right)}} \\
& \eqann{a} & 2 \mixing(\Delta) \sum_{q,r} \sqrt{(a_q c_r + b_q d_r)(a_q d_r+ b_q c_r)} \\
			      & = & 2 \mixing(\Delta) \sum_{q,r} \sqrt{(a_q^2 + b_q^2 )c_r d_r + (c_r^2 + d_r^2)a_q b_q } \\
			      & \eqann[\leq]{b} & 2 \mixing(\Delta) \sum_{q,r} \left( \sqrt{(a_q^2 + b_q^2 )c_r d_r} + \sqrt{(c_r^2 + d_r^2)a_q b_q } \right) \\
			      & \eqann[\leq]{c} & 2 \mixing(\Delta) \sum_{q,r} \left( \left(a_q + b_q \right)\sqrt{c_r d_r}  + \left(c_r+d_r\right)\sqrt{a_q b_q} \right) \\
			      & \eqann{d} &  \mixing(\Delta) \left( 2\sum_{r} \sqrt{c_r d_r}  + 2\sum_q \sqrt{a_q b_q} \right)\\
			      & \eqann{e} & 2 M(\Delta) Z_n(\Delta), 
	\end{IEEEeqnarray*}
	where~\eqannref{a} is by~\eqref{eq:abcd def} and~\eqref{eq:puvqr MDelta}; \eqannref{b} and \eqannref{c} follow from the inequality $\sqrt{\alpha + \beta} \leq \sqrt{\alpha} + \sqrt{\beta}$, which holds for $\alpha, \beta \geq 0$; \eqannref{d} holds since $a_q + b_q = \Prob{Q_i = q}$, $c_r+d_r = \Prob{R_i = r}$ and we sum over all $q$ and $r$, respectively; finally, \eqannref{e} is by stationarity. 

	Next,
	\begin{IEEEeqnarray*}{rCl}
		Z_n^+(\Delta) & = & 2\sum_{q,r,t} \sqrt{\Prob{V_i = 0, U_i + V_i = t, Q_i = q, R_i = r| D(\Delta)}} \\
		\IEEEeqnarraymulticol{3}{r}{\cdot \sqrt{\Prob{V_i = 1, U_i + V_i = t, Q_i = q, R_i = r| D(\Delta)}}} \\
		              & = & 2\sum_{q,r,t} \sqrt{\Prob{U_i = t, V_i = 0,  Q_i = q, R_i = r| D(\Delta)}} \\
		\IEEEeqnarraymulticol{3}{r}{\cdot \sqrt{\Prob{U_i = 1+t, V_i = 1,  Q_i = q, R_i = r| D(\Delta)}}} \\
			      & = & 2\sum_{q,r,t} \sqrt{ p(t,0,q,r) p(1+t,1,q,r)} \\ 
			      &\eqann[\leq]{a}& 2 M(\Delta) \sum_{q,r,t}\sqrt{ p(t,q) p(1+t,q) p(0,r) p(1,r)}\\ 
			      &\eqann{b}& 2 M(\Delta) \sum_{q,r,t}\sqrt{ p(0,q) p(1,q) p(0,r) p(1,r)}\\ 
&\eqann{c}& 2 M(\Delta) \cdot 2 \sum_{q,r}\sqrt{ a_q b_q c_r d_r}\\ 
			      &=& M(\Delta)\cdot 2\sum_{q}\sqrt{a_q b_q} \cdot 2 \sum_r \sqrt{c_r d_r} \\ 
			      &\eqann{d}& M(\Delta) (Z_n(\Delta))^2,
	\end{IEEEeqnarray*}
	where \eqannref{a} is by \Cref{lemm:mixing}, \eqannref{b} and \eqannref{c} are since $t$ can be either $0$ or $1$, and~\eqannref{d} is by stationarity. 
\end{IEEEproof}

\begin{lemma} \label{lemm:Kn satisfies 167}
Fix $0 \leq \Delta  < \deriod$, $N = 2 ^n \geq \deriod$, and $1 \leq i \leq N$. Then,
\[
	\hat{K}_n^-(\Delta) \leq \deriod\cdot\mixing(\Delta) \cdot (\hat{K}_n(\Delta))^2 \quad \mbox{and} \quad \hat{K}_n^+(\Delta) \leq 2 \cdot \hat{K}_n(\Delta) .
\]
\end{lemma}
This is an extension of \cite[Proposition 12]{ShuvalTal:19.2p} to our scenario.

\begin{IEEEproof}
	Recalling~\eqref{eq:p s0 sN s2N}, we denote
	\begin{equation*} 
	 \lambda(s_N) = \pi(s_0, s_N, s_{2N}), 
 \end{equation*}
 where we have deliberately omitted the dependence on $s_0, s_{2N}$. By summing over $s_0$ or $s_{2N}$ in $\mathcal{S}(\Delta)$ we obtain 
 \begin{IEEEeqnarray}{rCl} \label{eq:sum of lambda}
	 \sum_{s_0\in \mathcal{S}(\Delta)} \lambda(s_N) &=& P_{S_N,S_{2N}}(s_N,s_{2N}), \IEEEyessubnumber \\ 
	 \sum_{s_{2N}\in \mathcal{S}(\Delta)} \lambda(s_N) &=& P_{S_0,S_{N}}(s_0,s_{N}). \IEEEyessubnumber
\end{IEEEeqnarray}

	For $(s_0,s_N, s_{2N}) \in \mathcal{S}_N^3(\Delta)$, we have by~\eqref{eq:mixing def} and the Markov property
	\begin{equation} \label{eq:lambda inequality}
		\lambda(s_N) \leq \deriod \cdot M(\Delta)P_{S_0,S_N}(s_0,s_N)P_{S_N,S_{2N}}(s_N,s_{2N}).
	\end{equation}
We further define for $s_0,s_N,s_{2N} \in \mathcal{S}_N^3(\Delta)$ and $i$ clear from the context the following shorthand
\begin{IEEEeqnarray*}{rCl}
	p_{s_0}^{s_N}(u,q) &=& P_{U_i,Q_i|S_0, S_N}(u,q|s_0,s_N) \\  
	p_{s_0}^{s_N}(q) &=& \sum_u p_{s_0}^{s_N}(u,q) \\ 
	p_{s_N}^{s_{2N}}(v,r) &=& P_{V_i,R_i|S_N,S_{2N}}(v,r|s_N,s_{2N}) \\ 
	p_{s_N}^{s_{2N}}(r) &=& \sum_v p_{s_N}^{s_{2N}}(v,r) \\  
	p_{s_0}^{s_{2N}}(u,v,q,r) & = & P_{U_i,V_i,Q_i,R_i|S_0, S_{2N}}(u, v,q, r | s_0 , s_{2N}) \\
\bar{p}_{s_0}^{s_{2N}}(t,v,q,r) & = & P_{U_i+V_i,V_i,Q_i,R_i|S_0, S_{2N}}(t, v,q, r | s_0 , s_{2N}) \\
\bar{p}_{s_0}^{s_{2N}}(t,q,r) & = & \sum_v \bar{p}_{s_0}^{s_{2N}}(t,v,q,r) \IEEEyesnumber \label{eq:ptqr shorthand} \\ 
K_{s_0}^{s_N}(U_i|Q_i) &=& \sum_q \left|p_{s_0}^{s_N}(0,q)-p_{s_0}^{s_N}(1,q) \right|\\  
K_{s_N}^{s_{2N}}(V_i|R_i) &=& \sum_r \left|p_{s_N}^{s_{2N}}(0,r)-p_{s_N}^{s_{2N}}(1,r) \right|.  
\end{IEEEeqnarray*}
Denote
\[
	\pi(s_0,*,s_{2N}) = P_{S_0,S_{2N}}(s_0,s_{2N}).
\]
We then have
\begin{IEEEeqnarray*}{rCl} 
	\IEEEeqnarraymulticol{3}{l}{\pi(s_0,*,s_{2N})p_{s_0}^{s_{2N}}(u,v,q,r)} \\
\quad &=& \pi(s_0) P_{S_{2N}|S_0}(s_{2N}|s_0) p_{s_0}^{s_{2N}}(u,v,q,r)\\
      &=& \pi(s_0) P_{U_i,V_i,Q_i,R_i,S_{2N}|S_0}(u,v,q,r,s_{2N}|s_0)  \\
      &=& \pi(s_0) \sum_{s_N \in \mathcal{S}(\Delta)} P_{U_i,V_i,Q_i,R_i,S_N,S_{2N}|S_0}(u,v,q,r,s_N,s_{2N}|s_0)  \\
      &=& \pi(s_0) \sum_{s_N \in \mathcal{S}(\Delta)} P_{U_i,Q_i,S_N|S_0}(u,q,s_N|s_0)  \\
      \IEEEeqnarraymulticol{3}{r}{\cdot P_{V_i,R_i,S_{2N}|S_N}(v,r,s_{2N}|s_N) \quad}  \\
      &=& \pi(s_0) \sum_{s_N \in \mathcal{S}(\Delta)} P_{S_N|S_0}(s_N| s_0 ) p_{s_0}^{s_N}(u,q)  \\
      \IEEEeqnarraymulticol{3}{r}{\cdot P_{S_{2N}|S_N}(s_{2N} | s_N ) p_{s_N}^{s_{2N}}(v,r) \quad} \\ 
      &=& \sum_{s_N \in \mathcal{S}(\Delta)} \lambda(s_N) p_{s_0}^{s_N}(u,q) p_{s_N}^{s_{2N}}(v,r), \IEEEyesnumber \label{eq:important probability equality}
\end{IEEEeqnarray*}
where the latter equality is since $\pi(s_0, s_N,s_{2N}) = \pi(s_0)P_{S_N|S_0}(s_N|s_0)P_{S_{2N}|S_N}(s_{2N}|s_N)$. 

Denote the set $\mathcal{S}_\star = \{s_0,s_{2N}: (s_0, s_{2N}) \in \mathcal{S}(\Delta)\}$. Then, 
\[
	\hat{K}_n^{-}(\Delta) = \sum_{\mathcal{S}_\star}  \sum_{q,r} \pi(s_0,*,s_{2N}) \left| \bar{p}_{s_0}^{s_{2N}} (0,q,r) - \bar{p}_{s_0}^{s_{2N}} (1,q,r)  \right|.
\]
We first bound the inner sum for some fixed $(s_0, s_{2N}) \in \mathcal{S}_\star$.
\begin{IEEEeqnarray*}{rCl}
\IEEEeqnarraymulticol{3}{l}{\sum_{q,r} \pi(s_0,*,s_{2N}) \left| \bar{p}_{s_0}^{s_{2N}} (0,q,r) - \bar{p}_{s_0}^{s_{2N}} (1,q,r)  \right|} \\
&=& \sum_{q,r} \left|\pi(s_0,*,s_{2N})  \bar{p}_{s_0}^{s_{2N}} (0,q,r) -\pi(s_0,*,s_{2N})  \bar{p}_{s_0}^{s_{2N}} (1,q,r)  \right| \\
&\eqann{a}& \sum_{q,r} \left| \sum_{s_N \in \mathcal{S}(\Delta)} \lambda(s_N) \sum_v p_{s_N}^{s_{2N}}(v,r) ( p_{s_0}^{s_N}(v,q) - p_{s_0}^{s_N}(v+1,q)) \right| \\
&\eqann[\leq]{b}& \sum_{\mathclap{\substack{q,r, \\ s_N \in \mathcal{S}(\Delta)}}} \lambda(s_N)\left|  \sum_v p_{s_N}^{s_{2N}}(v,r) ( p_{s_0}^{s_N}(v,q) - p_{s_0}^{s_N}(v+1,q)) \right| \\
&=& \sum_{\mathclap{\substack{q,r, \\ s_N \in \mathcal{S}(\Delta)}}} \lambda(s_N) \left| p_{s_0}^{s_N}(0,q) - p_{s_0}^{s_N}(1,q) \right| \cdot \left| p_{s_N}^{s_{2N}}(0,r) - p_{s_N}^{s_{2N}}(1,r) \right| \\ 
&=& \sum_{s_N \in \mathcal{S}(\Delta)} \lambda(b) K_{s_0}^{s_N}(U_i|Q_i)K_{s_N}^{s_{2N}}(V_i|R_i) \\ 
&\eqann[\leq]{c}& \deriod \cdot \mixing(\Delta) \sum_{s_N \in \mathcal{S}(\Delta)}P_{S_0,S_N}(s_0,s_N)K_{s_0}^{s_N}(U_i|Q_i) \\ 
\IEEEeqnarraymulticol{3}{r}{\cdot P_{S_N,S_{2N}}(s_N,s_{2N})K_{s_N}^{s_{2N}}(V_i|R_i)} \\ 
&\eqann[\leq]{d}& \deriod\cdot \mixing(\Delta) \sum_{s_N \in \mathcal{S}(\Delta)}P_{S_0,S_N}(s_0,s_N)K_{s_0}^{s_N}(U_i|Q_i) \\ 
\IEEEeqnarraymulticol{3}{r}{\cdot \sum_{s_N' \in \mathcal{S}(\Delta)}P_{S_N,S_{2N}}(s_N',s_{2N})K_{s_N}^{s_{2N}}(V_i|R_i),}
\end{IEEEeqnarray*}
where \eqannref{a} is by~\eqref{eq:ptqr shorthand} and~\eqref{eq:important probability equality}, \eqannref{b} is by the triangle inequality, \eqannref{c} is by~\eqref{eq:lambda inequality}, and \eqannref{d} is by adding nonnegative terms to the sum.  
Finally, by summing over $\mathcal{S}_\star$ we obtain
\[ 
	\hat{K}_n^{-}(\Delta) = \deriod\cdot \mixing(\Delta) \cdot (\hat{K}_n(\Delta))^2.
\]

Next,
\[
	\hat{K}_n^{+}(\Delta) = \sum_{\mathcal{S}_\star}  \sum_{\mathclap{t,q,r}} \pi(s_0,*,s_{2N}) \left| \bar{p}_{s_0}^{s_{2N}} (t,0,q,r)\! -\! \bar{p}_{s_0}^{s_{2N}} (t,1,q,r)  \right|.
\]
Similar to the above, we first bound the inner sum for some fixed $(s_0, s_{2N}) \in \mathcal{S}_\star$, 
\begin{IEEEeqnarray*}{rCl}
	\IEEEeqnarraymulticol{3}{l}{\sum_{\mathclap{t,q,r}} \pi(s_0,*,s_{2N}) \left| \bar{p}_{s_0}^{s_{2N}} (t,0,q,r) - \bar{p}_{s_0}^{s_{2N}} (t,1,q,r)  \right|} \\
	&=& \sum_{\mathclap{t,q,r}} \Big|P_{S_0,S_{2N}}(s_0,s_{2N})\bar{p}_{s_0}^{s_{2N}}(t,0,q,r) \\ 
	\IEEEeqnarraymulticol{3}{r}{- P_{S_0,S_{2N}}(s_0,s_{2N})\bar{p}_{s_0}^{s_{2N}}(t,1,q,r) \Big|} \\ 
	&\eqann{a}& \sum_{\mathclap{t,q,r}} \Big| \sum_{s_N \in \mathcal{S}(\Delta)}\lambda(s_N)\Big(p_{s_0}^{s_{N}}(t,q)p_{s_N}^{s_{2N}}(0,r) \\ 
	\IEEEeqnarraymulticol{3}{r}{- p_{s_0}^{s_{N}}(t+1,q)p_{s_N}^{s_{2N}}(1,r)\Big) \Big|} \\ 
	&\eqann{b}& \frac{1}{2} \sum_{\mathclap{t,q,r}} \Bigg| \sum_{s_N \in \mathcal{S}(\Delta)}\lambda(s_N)\Big(p_{s_0}^{s_{N}}(q)\big(p_{s_N}^{s_{2N}}(0,r) - p_{s_N}^{s_{2N}}(1,r)\big) \Big) \\ 
		\IEEEeqnarraymulticol{3}{r}{+\sum_{s_N \in \mathcal{S}(\Delta)} \lambda(s_N)\Big( p_{s_N}^{s_{2N}}(r)\big(p_{s_0}^{s_{N}}(0,q) - p_{s_0}^{s_{N}}(1,q)\big)\Big) \Bigg|} \\ 
	&\eqann[\leq]{c}& \sum_{\substack{q, \\ s_N \in \mathcal{S}(\Delta)}} \lambda(s_N)p_{s_0}^{s_{N}}(q)\Big( \sum_r \left|p_{s_N}^{s_{2N}}(0,r) - p_{s_N}^{s_{2N}}(1,r)\right|\Big) \\ 
	\IEEEeqnarraymulticol{3}{r}{+ \sum_{\substack{r, \\ s_N \in \mathcal{S}(\Delta)}} \lambda(s_N)p_{s_0}^{s_{N}}(r)\Big( \sum_q \left|p_{s_0}^{s_{N}}(0,q) - p_{s_0}^{s_{N}}(1,q)\right|\Big)} \\ 
	&=& \sum_{s_N \in \mathcal{S}(\Delta)}\lambda(s_N)K_{s_N}^{s_{2N}}(V_i|R_i) + \sum_{s_N \in \mathcal{S}(\Delta)}\lambda(s_N)K_{s_0}^{s_{N}}(U_i|Q_i),
\end{IEEEeqnarray*}
where~\eqannref{a} is by~\eqref{eq:important probability equality}, \eqannref{b} is by the following algebraic identity that holds for any four numbers $a,b,c,d$, 
\[ ab-cd = \frac{(a+c)(b-d)+(b+d)(a-c)}{2},\] 
and \eqannref{c} is by the triangle inequality and since $t$ is binary. 

It remains to sum over $\mathcal{S}_\star$. Recalling~\eqref{eq:sum of lambda}, stationarity implies that
\begin{IEEEeqnarray*}{rCl}
\sum_{\mathcal{S}_N^3(\Delta)} \lambda(s_{N}) K_{s_N}^{s_{2N}}(V_i|R_i) &=& \sum_{\mathclap{s_N, s_{2N} \in \mathcal{S}_N^2(\Delta)}} P_{S_N, S_{2N}}(s_N, s_{2N}) K_{s_N}^{s_{2N}} (V_i|R_i) \\ &=& \hat{K}_n(\Delta) \\ 
\sum_{\mathcal{S}_N^3(\Delta)} \lambda(s_{N}) K_{s_0}^{s_{N}}(U_i|Q_i) &=& \sum_{\mathclap{s_0, s_{N} \in \mathcal{S}_N^2(\Delta)}} P_{S_0, S_{N}}(s_0, s_{N}) K_{s_0}^{s_{N}} (U_i|Q_i) \\ &=& \hat{K}_n(\Delta). 
\end{IEEEeqnarray*}
Thus, we obtain 
\[ 
	\hat{K}_n^+(\Delta) \leq 2 \hat{K}_n(\Delta). 
\] 
This completes the proof. 
\end{IEEEproof}

The following proposition establishes fast polarization. Note that this is not yet \Cref{thm:main}, since that requires conditioning on a finer event than $D(\Delta)$.

\begin{proposition} \label{prop:fast polarization} Fix $0 < \beta < 1/2$ and $0 \leq \Delta < \deriod$. Then, for $N= 2^n$,
	\begin{IEEEeqnarray}{l}
		\lim_{n \to \infty} \frac{1}{N}\left|\left\{ i : Z(U_i|Q_i, D(\Delta) ) < 2^{-N^{\beta}} \right\} \right|  =  1 - \mathcal{H}_{X|Y} , \IEEEeqnarraynumspace \label{eq:Z fast polarization limit} \\ 
		\lim_{n \to \infty} \frac{1}{N}\left|\left\{ i : K(U_i|Q_i, S_0, S_N,  D(\Delta)) < 2^{-N^{\beta}} \right\} \right|  =  \mathcal{H}_{X|Y} . \IEEEeqnarraynumspace \label{eq:K fast polarization limit}
	\end{IEEEeqnarray}
\end{proposition}

\begin{IEEEproof}
	We prove fast  polarization using \cite[Proposition 49]{ShuvalTal:25aa}. We first concentrate on~\eqref{eq:Z fast polarization limit}. By \Cref{lemm:Zn satisfies 167}, the process $Z_n(\Delta)$ satisfies \cite[Equation 167]{ShuvalTal:25aa}, with  $\kappa = 2 \mixing(\Delta)$. This also holds if we offset our process to start at some $n'$ instead of $0$. As a shorthand, we write $Z_n$ in place of $Z_n(\Delta)$ in the sequel. 

	Fix $\delta > 0$. Thus, \cite[Proposition 49]{ShuvalTal:25aa} implies that there exist $\eta > 0$ and $n_0$ such that for $\beta' = (\beta+1/2)/2$,
	\[
		\Prob{Z_{n+n'} \leq 2^{-2^{n\beta'}} \; \mbox{for all $n \geq n_0$} \Big| Z_{n'} \leq \eta  } \geq 1- \delta/2.
	\]
	We now show that we can choose $n'$ such that
	\begin{equation}
		\label{eq:enough small Zn'}
		\Prob{Z_{n'} \leq \eta} > 1 - \mathcal{H}_{X|Y} - \delta/2 .
	\end{equation}
		This will imply that
	\[
		\Prob{Z_{n+n'} \leq 2^{-2^{n\beta'}} \; \mbox{for all $n \geq n_0$},\; Z_{n'} \leq \eta  } > 1 -\mathcal{H}_{X|Y} - \delta.
	\]
	Hence, 
	\[
		\Prob{Z_{n+n'} \leq 2^{-2^{n\beta'}} \; \mbox{for all $n \geq n_0 $} } > 1- \mathcal{H}_{X|Y} - \delta.
	\]
	This implies that
	\begin{equation}\label{eq:Z limit inequality}
		\lim_{n \to \infty} \frac{1}{N}\left|\left\{ i : Z(U_i|Q_i, D(\Delta) ) < 2^{-N^{\beta}} \right\} \right|  \geq  1 - \mathcal{H}_{X|Y} 
	\end{equation}
	since $\beta' > \beta$ and $\delta > 0$ is arbitrary. All that remains to prove \eqref{eq:Z fast polarization limit}, is to show that the above ``$\geq$'' is in fact an equality. We will come back to this point later.

	To prove \eqref{eq:enough small Zn'}, we call upon \Cref{prop:slow polarization} with $\epsilon = \sqrt{\eta}$.  By, \eqref{eqs:limit set D epsilon}, \eqref{eqs:one minus entropy rate}, and \eqref{eq:HXY as limit}, 
	\[
		\lim_{n \to \infty} \Prob{H_n(\Delta) < \sqrt{\eta}} = 1 - \mathcal{H}_{X|Y} .
	\]
	Thus, there exists $n'$ such that 
	\[
		\Prob{H_{n'}(\Delta) < \sqrt{\eta}} = 1 - \mathcal{H}_{X|Y} - \delta/2 .
	\]
	By \cite[Lemma 1]{ShuvalTal:19.2p}, $H_{n'}(\Delta) < \sqrt{\eta}$ implies $Z_{n'} < \eta$, which establishes \eqref{eq:enough small Zn'}.

We now concentrate on~\eqref{eq:K fast polarization limit}. By \Cref{lemm:Kn satisfies 167}, the process $\hat{K}_n(\Delta)$ satisfies \cite[Equation 167]{ShuvalTal:25aa}, with  $\kappa = \max\{\deriod \mixing(\Delta),2\}$. As before, this also holds if we offset our process to start at some $n'$ instead of $0$. As a shorthand, we write $\hat{K}_n$ in place of $\hat{K}_n(\Delta)$ in the sequel. 

	Fix $\delta > 0$. Thus, \cite[Proposition 49]{ShuvalTal:25aa} implies that there exist $\eta > 0$ and $n_0$ such that for $\beta' = (\beta+1/2)/2$,
	\[
		\Prob{\hat{K}_{n+n'} \leq 2^{-2^{n\beta'}} \; \mbox{for all $n \geq n_0$} \Big| \hat{K}_{n'} \leq \eta  } \geq 1- \delta/2.
	\]
	We now show that we can choose $n'$ such that
	\begin{equation}
		\label{eq:enough small Kn'}
		\Prob{\hat{K}_{n'} \leq \eta} > \mathcal{H}_{X|Y} - \delta/2 .
	\end{equation}
		This will imply that
	\[
		\Prob{\hat{K}_{n+n'} \leq 2^{-2^{n\beta'}} \; \mbox{for all $n \geq n_0$},\; \hat{K}_{n'} \leq \eta  } > \mathcal{H}_{X|Y} - \delta.
	\]
	Hence, 
	\[
		\Prob{\hat{K}_{n+n'} \leq 2^{-2^{n\beta'}} \; \mbox{for all $n \geq n_0 $} } > \mathcal{H}_{X|Y} - \delta.
	\]
	This implies that
	\begin{equation}\label{eq:K limit inequality}
		\lim_{n \to \infty} \frac{1}{N}\left|\left\{ i : \hat{K}(U_i|Q_i, S_0, S_N, D(\Delta) ) < 2^{-N^{\beta}} \right\} \right|  \geq  \mathcal{H}_{X|Y} 
	\end{equation}
	since $\beta' > \beta$ and $\delta > 0$ is arbitrary. All that remains to prove \eqref{eq:K fast polarization limit}, is to show that the above ``$\geq$'' is in fact an equality. We will come back to this point later.

	To prove \eqref{eq:enough small Kn'}, we call upon \Cref{prop:slow polarization} with $\epsilon = \sqrt{1-\eta^2}$.  By, 
	\eqref{eqs:limit set S0 SN D one minus epsilon}, \eqref{eqs:entropy rate}, and \eqref{eq:HXY as limit}, 
\[
		\lim_{n \to \infty} \Prob{H_n(\Delta) > \sqrt{1- \eta^2}} = \mathcal{H}_{X|Y} .
	\]
	Thus, there exists $n'$ such that 
	\[
		\Prob{H_{n'}(\Delta) > \sqrt{1-\eta^2}} = \mathcal{H}_{X|Y} - \delta/2 .
	\]
	By \cite[Lemma 1]{ShuvalTal:19.2p}, $H_{n'}(\Delta) > \sqrt{1-\eta^2}$ implies $\hat{K}_{n'} < \eta$, which establishes \eqref{eq:enough small Kn'}.

	To complete the proof, we now show that the weak inequalities in \eqref{eq:Z limit inequality} and \eqref{eq:K limit inequality} are in fact equalities. To see this, first note by \cite[Lemma 2]{ShuvalTal:19.2p} that $Z(U_i|Q_i, S_0, S_N, D(\Delta)) \leq Z(U_i|Q_i, D(\Delta))$. Hence, \eqref{eq:Z limit inequality} implies that
	\begin{IEEEeqnarray}{l}
		\label{eq:hat Z limit inequality}
		\lim_{n \to \infty}\! \frac{1}{N}\!\left|\left\{ i : Z(U_i|Q_i, S_0, S_N, D(\Delta) ) < 2^{-N^{\beta}} \right\} \right| \! \geq\!  1 - \mathcal{H}_{X|Y} , \IEEEeqnarraynumspace
	\end{IEEEeqnarray}
	and a strict inequality in \eqref{eq:Z limit inequality} implies a strict inequality in \eqref{eq:hat Z limit inequality}. Now, assume to the contrary a strict inequality in either \eqref{eq:Z limit inequality} or \eqref{eq:K limit inequality}, or both. This implies a strict inequality in either \eqref{eq:hat Z limit inequality} or \eqref{eq:K limit inequality}, or both. Thus, for all $N$ large enough, there must exist an index $0 \leq i < N$, such that
	\begin{IEEEeqnarray*}{rCl}
		Z(U_i|Q_i, S_0, S_N, D(\Delta) ) &<& 2^{-N^{\beta}} , \\
		K(U_i|Q_i, S_0, S_N, D(\Delta) ) &<& 2^{-N^{\beta}} . 
	\end{IEEEeqnarray*}
	This cannot hold, since $K + Z \geq 1$, by \cite[Lemma 1]{ShuvalTal:19.2p}.
\end{IEEEproof}

\section{Proof of \Cref{thm:main}}
The following technical lemma will be used in our proof of \Cref{thm:main}. Its proof is given in Appendix~\ref{app:auxiliary proofs}.
\begin{lemma} \label{lemm:ZKH events}
Let $U$ be a binary random variable; $W$, $Q$ general random variables; and $A_1,A_2,\ldots,A_{\ell}$ mutually exclusive events. Denote $C = A_1 \cup A_2 \cup \cdots \cup A_{\ell}$. Then,
	\begin{IEEEeqnarray}{rCl}
		Z(U|Q,C) &\geq& \sum_{j=1}^{\ell} \Prob{A_j|C} Z(U|Q,A_j) , \label{eq:Z C inequality} \\
		K(U|Q,C) &\leq& \sum_{j=1}^{\ell} \Prob{A_j|C} K(U|Q,A_j) , \label{eq:K C inequality} \\
		H(W|Q,C) &\geq& \sum_{j=1}^{\ell} \Prob{A_j|C} H(W|Q,A_j) . \label{eq:H C inequality}
	\end{IEEEeqnarray}
\end{lemma}

\begin{IEEEproof}[Proof of \Cref{thm:main}]
Recall the factorization \eqref{eq:period factorization}, $\period = \deriod \cdot \qeriod$, where $\qeriod$ is odd and $\deriod$ is a power of $2$. Recall that $\varphi$ is the common phase of all the vertices in $\Psi_0$ and denote $\Delta = \varphi \bmod \deriod$. Next, note that by \eqref{eq:SN2} that for $N = 2^n \geq \deriod$ we have for all $(s_0,s_N) \in \Psi_0 \times \Psi_N$ that $(s_0,s_N) \in \mathcal{S}_N^2(\Delta)$.

	\Cref{prop:fast polarization} with $\beta' = (\beta+1/2)/2$ in place of $\beta$ implies that
	\begin{IEEEeqnarray*}{rCl}
	\lim_{n \to \infty} \frac{1}{N}\left|\left\{ i : Z(U_i|U_1^{i-1},Y_1^N, D(\Delta) ) < 2^{-N^{\beta}} \right\} \right| & = & 1 - \mathcal{H}_{X|Y} , \\
	\lim_{n \to \infty} \frac{1}{N}\left|\left\{ i : K(U_i|U_1^{i-1}, S_0, S_N, D(\Delta)) < 2^{-N^{\beta}} \right\} \right| & = & \mathcal{H}_{X} .
	\end{IEEEeqnarray*}
	Note that the second equality is obtained by considering the case in which the $Y$ are constant, and hence nothing is gained by conditioning on them.
The proof of \Cref{thm:main} will follow by showing that there is some $\xi > 0$ independent of $N$ and $0 \leq i < N$ such that
	\begin{equation}
		\label{eq:Z A bounded by Z}
		Z(U_i|U_1^{i-1},Y_1^N, A ) \leq \xi \cdot Z(U_i|U_1^{i-1},Y_1^N, D(\Delta) )
	\end{equation}
	and
	\begin{equation}
		\label{eq:K A bounded by hat K}
		K(U_i|U_1^{i-1}, A ) \leq \xi \cdot K(U_i|U_1^{i-1}, S_0, S_N, D(\Delta) ) ,
	\end{equation}
	for $N = 2^n$ large enough. Indeed, we claim that $\xi = \frac{1}{\mu \cdot \deriod}$ suffices for $N \geq N_0$, where $\mu$ and $N_0$ are as promised by \Cref{lemm:positive probability of a triplet}.

	For compactness, denote $\Psi_{0,N} \triangleq \Psi_0 \times \Psi_N$. To show \eqref{eq:Z A bounded by Z} we apply \Cref{lemm:ZKH events} with
	\begin{IEEEeqnarray*}{rCl}
		C &=& \{ (S_0, S_N) \in \mathcal{S}_N^2(\Delta) \} , \\
		A_1 &=& \{ (S_0, S_N) \in \Psi_{0,N} \} , \\
		A_2 &=& \{ (S_0, S_N) \in \mathcal{S}_N^2(\Delta) \setminus \Psi_{0,N} \} .
	\end{IEEEeqnarray*}
	Note that by definition $C$ and $A_1$ are simply $D(\Delta)$ and $A$, respectively. Thus,
	\begin{IEEEeqnarray*}{rCl}
		\IEEEeqnarraymulticol{3}{l}{Z(U_i|U_1^{i-1}, Y_1^N, D(\Delta))} \\
		\quad & \eqann[\geq]{a} & \Prob{A_1 | D(\Delta) } Z(U_i|U_1^{i-1}, Y_1^N, A_1) \\
		      & & {} + \Prob{A_2 | D(\Delta) } Z(U_i|U_1^{i-1}, Y_1^N, A_2) \\
		      & \eqann[\geq]{b} & \Prob{A | D(\Delta) } Z(U_i|U_1^{i-1}, Y_1^N, A) \\
		      & = & \Prob{(S_0,S_N) \in \Psi_{0,N} |D(\Delta) } \cdot Z(U_i|U_1^{i-1}, Y_1^N, A) \\ 
		      & = & \frac{\Prob{(S_0,S_N) \in \Psi_{0,N}}}{\Prob{D(\Delta)}} Z(U_i|U_1^{i-1}, Y_1^N, A) \\ 
		      &\eqann[\geq]{c}& \mu \cdot \deriod \cdot Z(U_i|U_1^{i-1}, Y_1^N, A) \\ 
		      &=& \xi^{-1} \cdot Z(U_i|U_1^{i-1}, Y_1^N, A)
	\end{IEEEeqnarray*}
	where~\eqannref{a} is by~\eqref{eq:Z C inequality}, \eqannref{b} is since $A_1 = A$ and probabilities are nonnegative, and \eqannref{c} is by \eqref{eq:prob of DDelta} and \Cref{lemm:positive probability of a triplet}. This yields~\eqref{eq:Z A bounded by Z}.

To show \eqref{eq:K A bounded by hat K}, we apply \Cref{lemm:ZKH events} with 
	\begin{IEEEeqnarray*}{rCl}
		C &=& \{ (S_0, S_N) \in \Psi_{0,N} \} , \\
		A_{s_0,s_N} &=& \{ (S_0, S_N) = (s_0,s_N) \} , 
	\end{IEEEeqnarray*}
	where in the above $(s_0,s_N)$ range over $\Psi_{0,N}$. Note that in this case, $C = A$, and clearly $C = \cup_{(s_0,s_N) \in \Psi_{0,N}} A_{s_0,s_N}$. Thus, 
	\begin{IEEEeqnarray*}{rCl}
		\IEEEeqnarraymulticol{3}{l}{K(U_i|U_1^{i-1}, A )} \\ 
		\quad &=& K(U_i|U_1^{i-1},C)) \\ 
	\quad &\eqann[\leq]{a}& \sum_{s_0, s_N} \Prob{A_{s_0,s_N}|C} K(U_i | U_1^{i-1}, S_0\!=\!s_0, S_N\!=\!s_N) \\ 
	\quad &\eqann{b}& \sum_{s_0, s_N} \frac{\Prob{A_{s_0,s_N}}}{\Prob{C}} K(U_i | U_1^{i-1}, S_0\!=\!s_0, S_N\!=\!s_N) \\ 
	\quad &\eqann{c}& \sum_{s_0, s_N} \frac{\Prob{A_{s_0,s_N},D(\Delta)}}{\Prob{C}} K(U_i | U_1^{i-1}, S_0\!=\!s_0, S_N\!=\!s_N) \\ 
	\quad &=& \sum_{s_0, s_N} \frac{\Prob{A_{s_0,s_N}|D(\Delta)}}{\Prob{D(\Delta)}^{-1}\Prob{C}} K(U_i | U_1^{i-1}, S_0\!=\!s_0, S_N\!=\!s_N) \\ 
	\quad &\eqann[\leq]{d}& \frac{1}{\mu \cdot \deriod} \sum_{s_0, s_N} \Prob{A_{s_0,s_N}|D(\Delta)} K(U_i | U_1^{i-1}, S_0\!=\!s_0, S_N\!=\!s_N), \\
	\quad & = & \xi \cdot K(U_i | U_1^{i-1}, S_0, S_N , D(\Delta)), 
	\end{IEEEeqnarray*}
	where \eqannref{a} is by~\eqref{eq:K C inequality}, \eqannref{b} is since $A_{s_0,s_N} \subseteq C$, \eqannref{c} is since $A_{s_0, s_N} \subseteq D(\Delta)$, and \eqannref{d} is by~\eqref{eq:prob of DDelta} and \Cref{lemm:positive probability of a triplet}. This yields \eqref{eq:K A bounded by hat K}, completing the proof. 
\end{IEEEproof}

\begin{appendices} \section{Auxiliary Proofs} \label{app:auxiliary proofs}
	\begin{IEEEproof}[Proof of \Cref{lemm:positive probability of a triplet}] 
	By the Markov property, we must show that there exist an $N_0$ and a $\mu$ such that for all $N \geq N_0$ and all $s_0,s_N,s_{2N}$ satisfying \eqref{eq:correct phase differences},
	\begin{multline*}
		\pi(s_0,s_N,s_{2N}) = \\
		\pi(s_0) \Prob{S_N = s_N | S_0 = s_0} \Prob{S_{2N} = s_{2N} | S_N = s_N } > \mu .
	\end{multline*}

	By irreducibility and stationarity, we have for all $s_0$ that
	\[
		\pi(s_0) \geq \pi_{\mathrm{min}} = \min_{s \in \mathcal{S}} \pi(s) > 0 .
	\]
	Hence, our claim will follow by showing that there exists $N_0$ and $\mu' > 0$ such that for all $N \geq N_0$ and all $s_0, s_N$ satisfying $\phi(s_N) - \phi(s_0) \equiv N \pmod{\period}$,
		\[
	\Prob{S_N = s_N | S_0 = s_0} \geq \mu' .
\]
	Namely, we can take $\mu = \pi_{\mathrm{min}} \cdot (\mu')^2/2$.

	By \cite[Theorem 1.3]{Seneta:81b}, there exists $N_0$ such that for all $\sigma, \omega \in \mathcal{S}$ such that $\phi(\omega) - \phi(\sigma) \equiv N_0 \pmod \period$, we have $\Prob{S_{N_0} = \omega | S_0 = \sigma} > 0$. We take $\mu'$ as the minimum of this probability over all $\sigma$ and $\omega$ as above. Now, for $N \geq N_0$ define $\mathcal{S}' = \{ \sigma \in \mathcal{S} :  \phi(\sigma) - \phi(s_0) \equiv N-N_0 \pmod{\period} \}$. We have
\begin{IEEEeqnarray*}{rCl}
	\IEEEeqnarraymulticol{3}{l}{\Prob{S_N = s_N | S_0 = s_0}} \\
	\quad &=& \sum_{\sigma \in \mathcal{S}'} \Prob{S_{N-N_0} = \sigma | S_0 = s_0 } \Prob{S_N = s_N | S_{N-N_0} = \sigma } \\
	      &\geq& \sum_{\sigma \in \mathcal{S}'} \Prob{S_{N-N_0} = \sigma | S_0 = s_0 } \mu' \\
	      &=& \mu' .
\end{IEEEeqnarray*}
The equalities follow by periodicity of the Markov chain and the definitions of $\mathcal{S}'$ and $\phi$; the inequality  follows since $\phi(s_N) - \phi(\sigma) \equiv N_0 \pmod{\period}$ for any $\sigma \in \mathcal{S}'$, by our definition of $\mathcal{S}'$, and our assumption that $\phi(s_N) - \phi(s_0) \equiv N \pmod{\period}$.
\end{IEEEproof}
	\begin{IEEEproof}[Proof of~\Cref{lemm:mixing}]
	For $(x_1^N,y_1^N), (x_{N+1}^{2N}, y_{N+1}^{2N}) \in \mathcal{X}^N \times \mathcal{Y}^N$, denote by $a(x_1^N,y_1^N)$ and $b(x_{N+1}^{2N}, y_{N+1}^{2N})$ the following events:
\begin{IEEEeqnarray*}{rCl}
	a(x_1^N,y_1^N) & = & \left\{(X_1^N,Y_1^N) =  (x_1^N,y_1^N)\right\}, \\
	b(x_{N+1}^{2N}, y_{N+1}^{2N}) & = & \left\{(X_{N+1}^{2N}, Y_{N+1}^{2N}) = (x_{N+1}^{2N}, y_{N+1}^{2N})\right\}.
\end{IEEEeqnarray*}
We start by noting the following: it suffices to prove that for all $(x_1^N,y_1^N), (x_{N+1}^{2N}, y_{N+1}^{2N}) \in \mathcal{X}^N \times \mathcal{Y}^N$ we have
	\begin{multline}
		\label{eq:mixingab}
		\Prob{a(x_1^N,y_1^N), b(x_{N+1}^{2N}, y_{N+1}^{2N})  \Big| D(\Delta) } \leq \mixing(\Delta) \\
		\cdot \Prob{ a(x_1^N,y_1^N)\Big| D(\Delta) )} \cdot \Prob{b(x_{N+1}^{2N}, y_{N+1}^{2N}) \Big| D(\Delta) } .
	\end{multline}
	Indeed, this would imply \eqref{eq:mixingAB}, since
	\begin{IEEEeqnarray*}{rCl}
		\IEEEeqnarraymulticol{3}{l}{\Prob{(X_1^N,Y_1^N) \in A \quad \mbox{and} \quad (X_{N+1}^{2N}, Y_{N+1}^{2N}) \in B \Big| D(\Delta) }} \\
		& = & \sum_{(x_1^N,y_1^N) \in A} \sum_{ (x_{N+1}^{2N}, y_{N+1}^{2N}) \in B} \Prob{a(x_1^N,y_1^N), b(x_{N+1}^{2N}, y_{N+1}^{2N})  \Big| D(\Delta) } \\
		& \leq & \mixing(\Delta) \sum_{(x_1^N,y_1^N) \in A} \Prob{a(x_1^N,y_1^N)  | D(\Delta) }  \\
		\IEEEeqnarraymulticol{3}{r}{\cdot \sum_{ (x_{N+1}^{2N}, y_{N+1}^{2N}) \in B}  \Prob{b(x_{N+1}^{2N}, y_{N+1}^{2N})  \Big| D(\Delta) }} \\
		& = & \mixing(\Delta) \cdot \Prob{(X_1^N,Y_1^N) \in A \Big| D(\Delta) )} \\
		\IEEEeqnarraymulticol{3}{r}{\cdot \Prob{(X_{N+1}^{2N}, Y_{N+1}^{2N}) \in B \Big| D(\Delta) }.} 
	\end{IEEEeqnarray*}

	To prove \eqref{eq:mixingab}, we start with its LHS and note that
	\begin{IEEEeqnarray*}{rCl}
	\IEEEeqnarraymulticol{3}{l}{\Prob{a(x_1^N,y_1^N), b(x_{N+1}^{2N}, y_{N+1}^{2N})  \Big| D(\Delta) }} \\
& = & \sum_{s_N \in \mathcal{S}(\Delta)} \Prob{a(x_1^N,y_1^N), S_N = s_N, b(x_{N+1}^{2N}, y_{N+1}^{2N})  \Big| D(\Delta) } \\
&\eqann{a}& \sum_{s_N \in \mathcal{S}(\Delta)} \Prob{S_N = s_N | D(\Delta) } \cdot \Prob{a(x_1^N,y_1^N) \Big|S_N = s_N, D(\Delta) } \\
\IEEEeqnarraymulticol{3}{r}{\cdot \Prob{b(x_{N+1}^{2N}, y_{N+1}^{2N})  \Big|S_N = s_N, D(\Delta) }} \\
&\eqann{b}& \sum_{s_N \in \mathcal{S}(\Delta)} \Prob{S_N = s_N | D(\Delta) } \cdot \frac{\Prob{a(x_1^N,y_1^N) ,S_N = s_N\Big| D(\Delta) }}{\Prob{S_N = s_N \Big| D(\Delta) }} \\
\IEEEeqnarraymulticol{3}{r}{\cdot \frac{\Prob{b(x_{N+1}^{2N}, y_{N+1}^{2N}) , S_N = s_N\Big| D(\Delta) }}{\Prob{S_N = s_N \Big| D(\Delta) }}} \\
&=& \sum_{s_N \in \mathcal{S}(\Delta)} \frac{1}{\Prob{S_N = s_N | D(\Delta) }} \cdot \Prob{a(x_1^N,y_1^N) ,S_N = s_N\Big| D(\Delta) } \\
\IEEEeqnarraymulticol{3}{r}{\cdot \Prob{b(x_{N+1}^{2N}, y_{N+1}^{2N}) , S_N = s_N\Big| D(\Delta) }} \\
&\eqann{c}& \sum_{s_N \in \mathcal{S}(\Delta)} \frac{1/\deriod}{\pi_{s_N}} \cdot \Prob{a(x_1^N,y_1^N) ,S_N = s_N\Big| D(\Delta) } \\
\IEEEeqnarraymulticol{3}{r}{\cdot \Prob{b(x_{N+1}^{2N}, y_{N+1}^{2N}) , S_N = s_N\Big| D(\Delta) }} \\
  & \eqann[\leq]{d} & \mixing({\Delta})\sum_{s_N \in \mathcal{S}(\Delta)}  \cdot \Prob{a(x_1^N,y_1^N) ,S_N = s_N\Big| D(\Delta) } \\
\IEEEeqnarraymulticol{3}{r}{\cdot \Prob{b(x_{N+1}^{2N}, y_{N+1}^{2N}) , S_N = s_N\Big| D(\Delta) }} \\
  & \eqann[\leq]{e} & \mixing({\Delta})\sum_{s_N \in \mathcal{S}(\Delta)} \sum_{s_N' \in \mathcal{S}(\Delta)} \Prob{a(x_1^N,y_1^N) ,S_N = s_N\Big| D(\Delta) } \\
\IEEEeqnarraymulticol{3}{r}{\cdot \Prob{b(x_{N+1}^{2N}, y_{N+1}^{2N}) , S_N = s_N'\Big| D(\Delta) }} \\
& = & \mixing({\Delta}) \sum_{s_N \in \mathcal{S}(\Delta)}\Prob{a(x_1^N,y_1^N) ,S_N = s_N\Big| D(\Delta) } \\
  \IEEEeqnarraymulticol{3}{r}{\cdot \sum_{s_N' \in \mathcal{S}(\Delta)} \Prob{b(x_{N+1}^{2N}, y_{N+1}^{2N}) , S_N = s_N'\Big| D(\Delta) }} \\
& = & \mixing({\Delta}) \cdot \Prob{a(x_1^N,y_1^N) \Big| D(\Delta) } \cdot \Prob{b(x_{N+1}^{2N}, y_{N+1}^{2N}) \Big| D(\Delta) } ,
	\end{IEEEeqnarray*}
	where \eqannref{a} is by the Markov property; \eqannref{b} is valid since the denominator is positive by irreducibility and stationarity;
	\eqannref{c} is since for $s_N \in \mathcal{S}_N$, we have
	\[ 
	\frac{1}{\Prob{S_N = s_N | D(\Delta)}} = \frac{\Prob{D(\Delta)}}{\Prob{S_N=s_N, D(\Delta)}} 
					       = \frac{1/\deriod}{\pi_{s_N}}, 
					       \] 
					       where the latter equality holds since $D(\Delta) \subseteq \{S_N = s_N\}$ and by~\eqref{eq:prob of DDelta}; \eqannref{d} is by \eqref{eq:mixing def}; and \eqannref{e} is since the sum over $s_N'$ adds nonnegative terms to the RHS.
This completes the proof.				    \end{IEEEproof}

\begin{IEEEproof}[Proof of \Cref{lemm:ZKH events}]
	Since the events $(A_j)_j$ are mutually exclusive and their union is $C$, we have $\Prob{A_j|C} = \Prob{A_j}/\Prob{C}$. 
	To keep the notation light, for a general event $A$, we denote $\Prob{U = u, Q = q | A} = p(u,q|A)$ and $\Prob{U=u,Q=q,A} = p(u,q,A)$. Observe that $p(u,q|A) =  p(u,q,A)/\Prob{A}$. 

	Recall that by the Cauchy-Schwarz inequality, for sequences $(a_j)_j$ and $(b_j)_j$ of the same length we have 
	\[ 
		\sqrt{\left(\sum_j a_j^2\right) \left(\sum_{j'} b_{j'}^2\right)} \geq \sum_j a_j b_j .
	\]
	Using the above inequality with $a_j = \sqrt{p(0,q,A_j)}$ and $b_{j} = \sqrt{p(1,q,A_j)}$, we obtain
	\begin{IEEEeqnarray*}{rCl}
		Z(U|Q,C) &=& 2 \sum_q \sqrt{p(0,q|C) p(1,q|C)} \\
		         &=& \frac{2}{\Prob{C}} \sum_q \sqrt{p(0,q,C) p(1,q,C)} \\
			 &=& \frac{2}{\Prob{C}} \sum_q \sqrt{ \left( \sum_{j=1}^\ell p(0,q,A_j) \right)\left( \sum_{j'=1}^\ell p(1,q,A_{j'}) \right) } \\
			 &\geq& \frac{2}{\Prob{C}} \sum_q \sum_{j=1}^\ell \sqrt{ p(0,q,A_j) \cdot p(1,q,A_j)} \\
			 &=& 2  \sum_{j=1}^\ell \frac{\Prob{A_j}}{\Prob{C}}  \sum_q \sqrt{ p(0,q|A_j) \cdot p(1,q|A_j)} \\
			 &=& \sum_{j=1}^\ell \Prob{A_j|C} Z(U|Q,A_j) .
	\end{IEEEeqnarray*}
	This yields \eqref{eq:Z C inequality}.

To prove \eqref{eq:K C inequality}, we have with the above notation and using the triangle inequality,
\begin{IEEEeqnarray*}{rCl}
		K(U|Q,C) &=&  \sum_q \left|p(0,q|C)  - p(1,q|C)\right| \\
			 &=& \frac{1}{\Prob{C}} \sum_q \left|p(0,q,C)  - p(1,q,C)\right| \\
			 &=& \frac{1}{\Prob{C}} \sum_q \left| \sum_{j=1}^\ell \left( p(0,q,A_j)  - p(1,q,A_j) \right) \right| \\
			 &\leq& \frac{1}{\Prob{C}} \sum_q \sum_{j=1}^\ell \left|  p(0,q,A_j)  - p(1,q,A_j) \right| \\
			 &=& \sum_{j=1}^\ell \frac{1}{\Prob{C}}  \sum_q   \left|  p(0,q,A_j)  - p(1,q,A_j) \right| \\
			 &=& \sum_{j=1}^\ell \frac{\Prob{A_j}}{\Prob{C}} \sum_q   \left|  p(0,q|A_j)  - p(1,q|A_j) \right| \\
			 &=& \sum_{j=1}^\ell \Prob{A_j|C} K(U|Q,A_j) .
\end{IEEEeqnarray*}

To prove \eqref{eq:H C inequality}, we define an indicator random variable $J$ that under $C$ equals $j$ if we are under event $A_j$. Since conditioning reduces entropy,
\begin{IEEEeqnarray*}{rCl}
	H(W|Q,C) \geq H(W|Q,C,J) &=& \sum_{j=1}^\ell \Prob{A_j|C} H(W|Q,C,J=j) \\
				 &=& \sum_{j=1}^\ell \Prob{A_j|C} H(W|Q,A_j) .
	\end{IEEEeqnarray*}
This completes the proof.
\end{IEEEproof}
\end{appendices}

\twobibs{
\bibliographystyle{IEEEtran} 
\bibliography{mybib.bib} 
}
{
\ifdefined\bibstar\else\newcommand{\bibstar}[1]{}\fi

}

\end{document}